\documentstyle[preprint,aps,epsf]{revtex}
\tighten

\begin{document}

\draft

\title{A Constrained Path Monte Carlo Method \\ 
for Fermion Ground States}

\author{Shiwei Zhang}
\address{Center for Nonlinear Studies and Theoretical Division\\
Los Alamos National Laboratory, Los Alamos, NM 87545\\
and\\
Department of Physics\\
The Ohio State University, Columbus, OH 43210}
\author{J.~Carlson and J.~E.~Gubernatis}
\address{Theoretical Division,
Los Alamos National Laboratory, Los Alamos, NM 87545}

\date{\today}
\maketitle

\begin{abstract}
We describe and discuss a recently proposed quantum Monte Carlo
algorithm to compute the ground-state properties of various systems of
interacting fermions. In this method, the ground state is projected
from an initial wave function by a branching random walk in an
over-complete basis of Slater determinants. By constraining the
determinants according to a trial wave function $|\psi_T \rangle$, we
remove the exponential decay of signal-to-noise ratio characteristic
of the sign problem. The method is variational and is exact if
$|\psi_T\rangle$ is exact. We illustrate the method by describing in
detail its implementation for the two-dimensional one-band Hubbard
model. We show results for lattice sizes up to $16\times 16$ and for
various electron fillings and interaction strengths.  Besides highly
accurate estimates of the ground-state energy, we find that the method
also yields reliable estimates of other ground-state observables, such
as superconducting pairing correlation functions.  We conclude by
discussing possible extensions of the algorithm.
\end{abstract}

\bigskip

\pacs{PACS numbers: 02.70.-c, 71.10.+x, 71.20.Ad, 71.45.Nt}


\section{INTRODUCTION}

We describe a new quantum Monte Carlo (QMC) algorithm to compute the
ground-state ($T=0\,$K) properties of systems of interacting fermions.
Our method, which is approximate, removes the {\it exponential\/}
scaling of computation time with system size that is characteristic of
the infamous fermion ``sign problem''\cite{sign1,sign2} in QMC
simulations \cite{GFMC,CepRMP,Bindernew}. Here we discuss the general
concepts of the algorithm and then describe details for its
implementation using the Hubbard model as an example. The test results
we present will show that the algorithm makes it possible to compute
accurately, in computation times scaling {\it algebraically} with
system size, general ground-state properties, such as superconducting
pairing correlation functions. A brief description of the basic CPMC
algorithm and some of the results on the Hubbard model were published
earlier\cite{PRL}. The algorithm, as it will be detailed here, can
also be directly applied to study many other lattice models of
electron correlations, such as the extended Hubbard model, the
Anderson lattice model, etc., where computer simulations with existing
QMC algorithms are often difficult and sometimes
impossible. Application of the method to more general problem classes,
such as atoms, molecules, and nuclei, is currently under study.

The new algorithm, called the constrained path Monte Carlo (CPMC)
method, has two main ingredients: The {\it first\/} is casting the
projection of the ground state from an arbitrary initial state as
importance-sampled branching random walks in a space of
Slater determinants. The {\it second\/} ingredient, used only to deal
with the sign problem, is constraining the paths of the random walks
so that any Slater determinant generated maintains a positive overlap
with a known trial wave function $|\psi_T\rangle$.

The first of the two ingredients is an exact procedure. As we will
illustrate, it combines important advantages of two
existing methods, the Green's function Monte Carlo (GFMC)
\cite{GFMC,MHK62,note1} and the auxiliary-field quantum Monte Carlo (AFQMC)
\cite{BSS,AFQMC,LGbook,note2} methods. For example, our method shares with 
the latter the ease of computing expectation values of certain
correlation functions, which are crucial to probe physical properties
but which are often hard to compute accurately by the standard GFMC
methods.  On the other hand, it shares the GFMC concept of importance
sampling with a trial wave function $|\psi_T\rangle$, which greatly
improves its efficiency over the AFQMC method. In addition, the
realization of the projection by open-ended random walks along the
imaginary-time direction, in contrast to the standard AFQMC
formulation, makes it practical and easy to implement the second
ingredient, the constrained path approximation, and hence to eliminate
the exponential scaling due to the sign problem.

The constrained path approximation ensures that the Monte Carlo
representation of the projected ground-state has {\it no} asymptotic
signal-to-noise ratio decay in imaginary time. The resulting method is
variational, with the computed ground-state energy being an upper
bound, and becomes exact if $|\psi_T\rangle$ is exact.  The
constrained-path approximation builds upon the positive projection
technique of Fahy and Hamann \cite{Fahy}, but can also be viewed as a
generalization of the fixed-node\cite{Anderson,Moskowitz,FN}
approximation in the GFMC method. Because of the different bases in
which the approximations are applied, the effect of the constrained
path approximation is expected to be {\it different\/} from that of
the fixed-node approximation.

In Section II, we will summarize the Green's function Monte Carlo and
the auxiliary-field quantum Monte Carlo methods for ground-state
calculations. Here, we will establish the necessary concepts and
formalisms from these existing approaches that are integral parts of
our new method. In Section III, we describe the CPMC method in general
terms, focusing on the concept of the importance-sampled random walks
in Slater-determinant space, the nature and consequences of the
constrained-path approximation, and the computation of expectation
values. Implementations issues are discussed in Section IV in the
context of the one-band Hubbard model. In Section V, we report results
for this model that illustrate the accuracy and performance
characteristics of the new method. Finally, in Section VI, we
summarize and discuss several simple extensions of the CPMC method.

\section{BACKGROUND}

In this section, we summarize the AFQMC method and also sketch a
particular GFMC method, namely the diffusion Monte Carlo (DMC)
method\cite{Anderson,Moskowitz}, that is most analogous to our new
method. In discussing the DMC method, the approach we use is not
standard, but rather it is one designed to provide the necessary
groundwork for the description of CPMC. Both the AFQMC and the GFMC
methods contain elements important to the CPMC method. For example,
the basic techniques of the AFQMC method, such as Hubbard-Stratonovich
transformation, imaginary-time propagation of Slater determinants, and
matrix multiplication stabilization
\cite{BSS,LGbook,white}, are shared by the CPMC method, on the other hand 
the random walk realization of the propagation\cite{MHK62,fwd},
importance sampling in the random walks by use of a known trial
function\cite{GFMC,MHK62}, and the fixed-node
approximation\cite{Anderson} are all GFMC concepts of much relevance.

Most ground-state quantum Monte Carlo methods are based on
\begin{equation}
  |\psi_0\rangle \propto \lim_{\tau\rightarrow\infty}
             e^{-\tau H}|\psi_T\rangle;
\end{equation}
that is, the ground state $|\psi_0\rangle$ can be projected from
any known trial state $|\psi_T\rangle$ that satisfies $\langle
\psi_T|\psi_0\rangle\ne0$.
In a numerical method, the limit can be obtained iteratively by
\begin{equation}
  |\psi^{(n+1)}\rangle = e^{-\Delta\tau H}|\psi^{(n)}\rangle,
\label{eq:process}
\end{equation}
where $|\psi^{(0)}\rangle = |\psi_T\rangle$.  With a small
$\Delta\tau$, the first-order Trotter approximation can be used:
\begin{equation}
   e^{-\Delta\tau H} \approx e^{-\Delta\tau K}e^{-\Delta\tau V}.
\label{eq:trotter}
\end{equation}
Typically, $K$ and $V$ are the kinetic and potential energy operators. More
generally, they are the one- and two-body interaction operators.

\subsection{Auxiliary-field quantum Monte Carlo method}

In the AFQMC method, the operators and wave function are in a second
quantized representation, defined in terms of fermion creation and
destruction operators $c^\dagger$ and $c$. The basis is one of Slater
determinants:
\begin{equation}
    |\phi\rangle \equiv \phi_1^\dagger \phi_2^\dagger \cdots
                   \phi_{N_\sigma}^\dagger|0\rangle
\label{eq:slater}
\end{equation}
where 
\begin{equation}
  \phi_i^\dagger \equiv \sum_j c_j^\dagger\, \Phi_{ji}.
\label{eq:orbitals}
\end{equation}
$\Phi_{ji}$ are the elements of a matrix $\Phi$ of dimension $N\times
N_\sigma$, where $N$ is the size of the basis and $N_\sigma$ is the
number of fermions with spin $\sigma$. Each column of the matrix
$\Phi$ represents a single-particle orbital that is completely
specified by a vector of dimension $N$. One example of such a Slater
determinant is the Hartree-Fock (HF) solution $|\phi_{\rm HF}\rangle
=\prod_\sigma |\phi_{\rm HF}^\sigma\rangle$, where each $|\phi_{\rm
HF}^\sigma\rangle$ is defined by a matrix $\Phi_{\rm HF}^\sigma$ whose
columns are the $N_\sigma$ lowest HF eigenstates. For any two real
non-orthogonal Slater determinants, $|\phi \rangle$ and
$|\phi^\prime\rangle$, it can be shown that their overlap integral
\begin{equation}
  \langle \phi|\phi^\prime\rangle = \det(\Phi^{\rm T}\Phi^\prime),
\label{eq:ovlp}
\end{equation}
and single-particle Green's function
\begin{equation}
   G_{ij} \equiv \frac{\langle\phi|c_i c_j^\dagger |\phi^\prime\rangle}
                  {\langle\phi|\phi^\prime\rangle}
          = \delta_{ij} 
          - [\Phi^\prime(\Phi^{\rm T}\Phi^\prime)^{-1}\Phi^{\rm T}]_{ij}.
\label{eq:G}
\end{equation}

Now we consider the projection (\ref{eq:process}) in this
Slater-determinant basis. The trial wave function $|\psi_T\rangle$ can
be a linear combination of determinants, but without loss of
generality, we assume it is a single determinant. A key point is that
the projection of any Slater determinant by any operator of the form
\begin{equation}
e^{\sum_{ij}c_i^\dagger M_{ij}c_j}
\label{eq:spo}
\end{equation}
simply leads to another Slater determinant, i.e.,
\begin{equation}
   e^{\sum_{ij}c_i^\dagger M_{ij}c_j}|\phi\rangle =
	 {\phi^\prime}_1^\dagger {\phi^\prime}_2^\dagger \cdots 
         {\phi^\prime}_{N_\sigma}^\dagger|0\rangle
         \equiv |\phi^\prime\rangle
\label{eq:expo}
\end{equation}
with ${\phi^\prime}_i^\dagger = \sum_j c_j^\dagger\,
\Phi^\prime_{ji}$ and $\Phi^\prime\equiv e^{-M}\Phi$. 
 
The $e^{-\Delta\tau K}$ part of (\ref{eq:trotter}) has the
single-particle form of (\ref{eq:spo}). The $e^{-\Delta\tau V}$ part,
however, represents two-body interactions with $V
=1/2\sum_{ijkl}V_{ijkl}c_i^\dagger c_j^\dagger c_l c_k$. Following
Hubbard, we rewrite $V$ as a quadratic form:
\begin{equation}
   V = \sum_\alpha \lambda_\alpha
\left(\sum_{ij}c_i^\dagger R_{ij}^\alpha c_j\right)^2
     \equiv \sum_\alpha \lambda_\alpha\rho_\alpha^2,
\end{equation}
where the parameters $\lambda_\alpha$ and the matrix $R^\alpha$ are
defined by the elements $V_{ijkl}$ and the number of $\alpha$ is at
most $N^2$ but is often much smaller.  With this quadratic form, a
Hubbard-Stratonovich (HS) transformation of the two-body part of
(\ref{eq:trotter}) yields:
\begin{equation}
   e^{-\Delta\tau\sum_\alpha \lambda_\alpha\rho_\alpha^2} =
       \prod_\alpha \int_{-\infty}^\infty dx_\alpha 
            \frac{e^{-\frac{1}{2} x_\alpha^2}}{\sqrt{2\pi}}
           \exp\Bigl(x_\alpha \sqrt{-\Delta\tau\lambda_\alpha}
              \sum_{ij}c_i^\dagger R^\alpha_{ij} c_j\Bigr),
\label{eq:HStrans}
\end{equation}
where $x_\alpha$ is an auxiliary-field variable. Denoting the 
collection of such variables by $\vec x$ and
defining $B(\vec x)={\rm exp}[-\Delta\tau\sum c_i^\dagger K_{ij}c_j]
\prod_\alpha {\rm exp}[x_\alpha \sqrt{-\Delta\tau\lambda_\alpha} 
\sum c_i^\dagger R^\alpha_{ij}c_j]$, we obtain: 
\begin{equation}
      e^{-\Delta\tau H}= \int d\vec x\,  P(\vec x) B(\vec x),
\label{eq:HS}
\end{equation}
where $P(\vec x)=\prod_\alpha (e^{-\frac{1}{2}
x_\alpha^2}/{\sqrt{2\pi}})$ is a probability density function and
$B(\vec x)$ has the desired single-particle form of
(\ref{eq:spo}). The essence of the HS transformation is the conversion
of an interacting system into many {\it non-interacting} ones living
in fluctuating external auxiliary-fields, and the summation over all
such auxiliary-field configurations recovers the correct many-body
interactions. We note that different forms of this transformation
exist and that they can affect the algorithm performance, possibly to
a large degree. These issues are, however, not of concern here, as we
will only be describing the general algorithm.

With (\ref{eq:process}) and (\ref{eq:HS}),
the ground-state expectation $\langle {\cal
O}\rangle$ of some observable ${\cal O}$ can be computed by
$\langle \psi^{(n)} |{\cal O}|\psi^{(n)}\rangle /
\langle \psi^{(n)} | \psi^{(n)}\rangle$. The denominator is 
\begin{eqnarray}
  \langle\psi^{(0)}| e^{-n\Delta\tau H}\,
                     e^{-n\Delta\tau H}|\psi^{(0)}\rangle
   &=& \int \langle \psi_T|
       \Bigl[\prod_{l=1}^{2n} 
d\vec x^{(l)} P(\vec x^{(l)})B(\vec x^{(l)})\Bigr]|\psi_T\rangle\\
   &=& \int \Bigl[\prod_l d\vec x^{(l)} P(\vec x^{(l)})\Bigr]
         \det\Bigl(\Psi_T^{\rm T} \prod_l{\bf B}(\vec x^{(l)})\Psi_T\Bigr)
\label{eq:afqmc}
\end{eqnarray}
where ${\bf B}(\vec x)$ is the $N\times N$ matrix associated with the
single-particle operator $B(\vec x)$ and equations (\ref{eq:ovlp}),
(\ref{eq:expo}), and (\ref{eq:HS}) have been applied. In the AFQMC
method\cite{BSS}, $n$ is fixed and the many-dimensional integral in
(\ref{eq:afqmc}) is evaluated by a Monte Carlo (MC) method like the
Metropolis algorithm. The MC process
samples configurations $\{\vec x^{(1)},\vec
x^{(2)},\dots,\vec x^{(2n)} \}$ of the auxiliary-fields distributed
according to the absolute 
value of the integrand.  

In the AFQMC method, the sign problem occurs because in general the
determinant in (\ref{eq:afqmc}) is not always positive. In fact, its
average sign approaches zero exponentially as $n$ (or $N$) is
increased\cite{sign1}. The integral then becomes vanishingly
small. Thus an {\it exponential\/} growth in computation time is
required in its evaluation, since the MC samples, drawn from the {\it
absolute value\/} of the integrand, become dominated by noise. This
problem has remained largely uncontrolled, preventing general
simulations at low temperatures or large system sizes.

One attempt to control the sign problem was the positive projection
approximation proposed by Fahy and Hamann \cite{Fahy}. They used a
known wave function $|\psi_c\rangle$ and imposed $2n$ conditions
\begin{eqnarray}
 \langle \psi_T| B(\vec x^{(1)})
         B(\vec x^{(2)})\cdots B(\vec x^{(l)})|\psi_c\rangle
    &>& 0, \quad l=1,2,\dots,n\\
 \langle \psi_c| B(\vec x^{(l)})
         B(\vec x^{(l+1)})\cdots B(\vec x^{(2n)})|\psi_T\rangle
    &>&0, \quad  l=2n, 2n-1,\dots,n
\end{eqnarray}
in sampling the auxiliary fields. The approximation is similar in
spirit to that of the fixed-node approximation in the GFMC
method. However, the constraint is {\it non-local\/} in imaginary
time, as any change in ${\vec x}^{(l)}$ affects the constraint
conditions at {\it all} times between $l$ and $n$. Thus all auxiliary
fields had to be updated simultaneously and only paths satisfying all
constraint equations were accepted. The approach is hence
computationally very intensive. In our CPMC method, we adopt the
Fahy-Hamann concept of a constraining state but implement the
constraint in the context of a random walk in the space of Slater
determinants, which makes the procedure practical and straightforward.

\subsection{Diffusion Monte Carlo method}

The DMC method\cite{qmc_book} executes the iteration in
(\ref{eq:process}) as random walks in configuration space. When the
fixed-point condition is reached, the random walks sample positions in
configuration space from a distribution that represents the
unknown {\it amplitude} of the ground-state wave function.

We denote the configuration basis by $|R\rangle$, where $R\equiv
\{ \vec r_1, \vec r_2, \dots, \vec r_{N_\sigma}\}$ is the electron 
coordinates in the continuous three-dimensional space. In this basis,
the potential energy propagator $e^{-\Delta\tau V}$ in
(\ref{eq:trotter}) is diagonal, but the kinetic energy propagator
$e^{-\Delta\tau K}$, where $K=-\frac{1}{2}\sum_i \nabla_i^2$, is not. In
order to write the latter in a more suitable form for a Monte Carlo
treatment, we invoke a Hubbard-Stratonovich (HS) transformation:
\begin{equation}
   e^{\frac{1}{2}\Delta\tau \nabla_i^2} = 
       \int d x_i \frac{e^{-\frac{1}{2} x_i^2}}{(2\pi)^{3/2}}
           e^{\sqrt{\Delta\tau}x_i \cdot \nabla_i}.
\label{eq:dmc1}
\end{equation}
Since $e^{\sqrt{\Delta\tau}x_i \cdot \nabla_i}|R\rangle$ displaces
$r_i$ in $|R\rangle$ by $\sqrt{ \Delta\tau} x_i$, the effect of
$e^{-\Delta\tau K}$ on any $|R\rangle$ can be viewed stochastically as
``diffusing'' it to $|R+\sqrt{\Delta\tau}\vec x\rangle$, where each
component $x_i$ of the auxiliary field $\vec x$ is drawn from the
normal distribution function $P(x_i)= e^{-\frac{1}{2}
x_i^2}/(2\pi)^{3/2}$.

The wave function $|\psi^{(n)}\rangle$ 
can be expressed in terms of the amplitudes
$\langle R|\psi^{(n)}\rangle\equiv \psi^{(n)}(R)$. In the random walk
realization of the iteration, $\psi^{(n)}(R)$ is
represented by a finite ensemble of configurations $\{ R^{(n)}_k
\}$. At each stage, the Monte Carlo method provides the stochastic
sampling of ${\vec x_k^{(n)}}$ and consequently the movement
$|R^{(n)}_k\rangle \rightarrow |R^{(n)}_k+\sqrt{ \Delta\tau} 
\vec x_k^{(n)} \rangle
\equiv |R^{(n+1)}_k\rangle$ for each configuration in the
ensemble. The factor $e^{-\Delta\tau V(R^{(n)}_k)}$ translates into a
weight (branching factor) for the configuration. As the iteration
approaches the fixed-point condition, the weighted distribution of
configurations represents $\psi_0(R) = \langle R|\psi_0\rangle$.

The sign problem in the DMC method has a somewhat different character
than the sign problem in the AFQMC method. The Pauli exclusion
principle requires that the fermion wave function $\psi_0(R)$ change
sign if the positions of two electrons with the same spin are
interchanged. Unlike the AFQMC method, the DMC method does not impose
the anti-symmetric property in the projection process. Without
additional mechanisms, the DMC method naturally produces points
distributed according to the lowest eigenstate of the diffusion
equation. This state is symmetric and bosonic-like. There has only
been limited success in attempts to construct exact algorithms that
yield asymptotically (in $n$) a non-vanishing, anti-symmetric Monte
Carlo signal\cite{inter,SignExact}.

The fixed-node method \cite{Anderson,Moskowitz,FN} is an approximate
scheme to prevent the convergence to the bosonic-like ground state.
Antisymmetry in $\psi_0(R)$ implies that there are equivalent regions
in configuration space that are separated by a nodal surface on which
$\psi_0(R)=0$. The exact nodal surface is in general unknown.  In the
fixed-node approximation, a trial nodal surface is assumed, based on a
known trial wave function $|\psi_T\rangle$. A solution which is
everywhere positive is then sought in the region $\psi_T(R) >0$ by
imposing the boundary condition that $\psi^{(n)}(R)$ vanishes at
$\psi_T(R)=0$. Unless $\psi_T(R)=0$ happens to be the correct node,
the resulting DMC solution for the ground state is approximate. The
ground-state energy obtained is an upper bound\cite{Moskowitz}.

An important feature of the DMC method is importance sampling.  This
technique is necessary to reduce the variance of the computed results
to acceptable levels.  For brevity we will not discuss this technique
here. Instead, we will postpone such a discussion until it is needed
to complete our description of the CPMC algorithm.

\section{CONSTRAINED PATH MONTE CARLO METHOD}

We now describe the CPMC algorithm. It uses the
Hubbard-Stratonovich-based formalism of the AFQMC method, but a Monte
Carlo sampling procedure similar to that of the DMC method.  The
iterative process (\ref{eq:process}) becomes an open-ended random walk
in {\it Slater determinant space\/}. Within the framework of this
random walk, we introduce importance sampling and the constrained path
approximation.

We remark that any antisymmetric wave function can be written as a linear 
combination of Slater determinants, i.e.,
\begin{equation}
|\psi\rangle = \sum_\phi \chi_{\psi}(\phi)|\phi\rangle,
\label{eq:grdwf}
\end{equation}
where the sum is over each member of the Slater determinant basis.  As
introduced in Section II, we will always use 
$|\psi\rangle$ to denote antisymmetric wave functions and
$|\phi\rangle$ to denote a single Slater determinant. Contrary to the
configuration space used in the DMC method, the
Slater-determinant basis space of $|\phi\rangle$ is {\it
non-orthogonal\/} and {\it over-complete\/}.

\subsection{Importance-sampled random walk formulation}

Using (\ref{eq:HS}), we write the iterative equation
(\ref{eq:process}) as
\begin{equation}
   |\psi^{(n+1)}\rangle = \int d\vec x\, P(\vec x) B(\vec x)|\psi^{(n)}\rangle.
\label{eq:cpmc}
\end{equation}
In the Monte Carlo realization of this iteration, we represent the
wave function at each stage by a finite ensemble of Slater
determinants, i.e.,
\begin{equation}
|\psi^{(n)}\rangle \propto \sum_k |\phi^{(n)}_k\rangle.
\label{eq:wf}
\end{equation}
Here $k$ labels the Slater determinants and an overall normalization
factor of the wave function has been omitted. The Slater determinants
are referred to as {\it random walkers\/} as they are generated by the
random walk. At any stage of the iteration, the sum will be over only
part of the basis as the determinants in this sum are statistical
samples whose distribution represents the linear coefficient
$\chi_{\psi^{(n)}}$ in (\ref{eq:grdwf}). The statistical accuracy of
this representation increases algebraically as the number of
independent samples is increased. In the remainder of the paper,
equation (\ref{eq:wf}) will serve as the definition of the Monte Carlo
representation of a wave function in the CPMC method. We will start
from an initial ensemble where $|\phi^{(0)}_k\rangle=|\psi_T\rangle$
\cite{notepsiT}.

One step of the iteration involves the propagation of each walker
according to (\ref{eq:cpmc}). Since the non-interacting operator
$B(\vec x)$ operating on any Slater determinant leads to another
Slater determinant, an analytical realization of this propagation for
each walker would yield a linear combination of many Slater
determinants. In our random walk, this propagation is achieved
stochastically by Monte Carlo (MC) sampling of $\vec x$:
\begin{equation}
|\phi^{(n+1)}_k\rangle 
\leftarrow 
\int d\vec x\,P(\vec x) B(\vec x) |\phi_k^{(n)}\rangle;
\label{eq:cpmcsampling}
\end{equation}
that is, for each random walker we choose an
auxiliary-field configuration $\vec x$ from the probability density
function $P(\vec x)$ and  propagate the walker to a new one via
$|\phi^{(n+1)}_k\rangle=B(\vec x)|\phi^{(n)}_k\rangle$. We repeat
this procedure for {\it all\/} walkers in the population.  These
operations accomplish one step of the random walk. The new population
represents $|\psi^{(n+1)}\rangle$ in the sense of
(\ref{eq:wf}).
These steps are iterated indefinitely. After an equilibration phase,
all walkers thereon are MC samples of the ground-state wave function
$|\psi_0\rangle$ and ground-state properties can be computed.

In order to improve the efficiency of (\ref{eq:cpmc}) and make it a
practical algorithm, an importance sampling scheme is required. In the
procedure just described, no information is contained in the sampling
of $\vec x$ on the importance of the resulting determinant in
representing $|\psi_0\rangle$, yet such information is clearly
important. For example, the ground-state energy is given by $E_0\equiv
\langle \psi_T |H|\psi_0\rangle/\langle \psi_T |
\psi_0\rangle$.  Hence, estimating $E_0$ requires estimating the denominator
by $\sum_\phi \langle\psi_T | \phi\rangle$, in which $|\phi\rangle$
denotes random walkers after equilibration.  Since these walkers are
sampled with no knowledge of $\langle \psi_T | \phi\rangle$, terms in
the summation over $\phi$ can have large fluctuations that lead to
large statistical errors in the MC estimate of the denominator,
thereby in that of $E_0$.

To introduce importance sampling, we iterate a modified equation with
a modified wave function, without changing the underlying eigenvalue
problem of (\ref{eq:cpmc}). Specifically, for each Slater determinant
$|\phi\rangle$, we define an importance function
\begin{equation}
O_T(\phi)\equiv \langle \psi_T|\phi\rangle, 
\label{eq:O_T}
\end{equation}
which estimates its overlap with the ground-state wave function. We can
then rewrite equation (\ref{eq:cpmc}) as
\begin{equation}
   |\tilde \psi^{(n+1)}\rangle = \int d\vec x \tilde P(\vec x) B(\vec x)|
      \tilde \psi^{(n)}\rangle,
\label{eq:impcpmc}
\end{equation}
where the modified wave function is
\begin{equation}
|\tilde \psi^{(n)}\rangle 
= \sum_\phi O_T(\phi) \chi_{\psi^{(n)}}(\phi)|\phi\rangle
\label{eq:modwf}
\end{equation}
and the modified ``probability density function'' is
\begin{equation}
\tilde P(\vec x)={O_T(\phi^{(n+1)}) 
            \over O_T(\phi^{(n)})} P(\vec x).
\label{eq:impcpmc_p}
\end{equation}
We note that $\tilde P(\vec x)$ is a function of both the future 
$|\phi^{(n+1)}\rangle$ and 
the current $|\phi^{(n)}\rangle$ positions in Slater-determinant space.
It is trivially verified that equations 
 (\ref{eq:cpmc}) and (\ref{eq:impcpmc}) are identical.

In the random walk, the ensemble of walkers
$\{\,|\phi^{(n)}_k\rangle\,\}$ now represents the modified wave
function $|\tilde \psi^{(n)}\rangle$, which is to say that their
distribution represents the function $\chi_{\psi^{(n)}}O_T$.  The
iterative relation for each walker is again given by
(\ref{eq:cpmcsampling}), but with $P(\vec x)$ replaced by $\tilde
P(\vec x)$.  The latter is in general not a normalized probability
density function, and we denote the normalization constant for walker
$k$ by $N(\phi_k^{(n)})$ and rewrite (\ref{eq:cpmcsampling}) as
\begin{equation}
|\phi^{(n+1)}_k\rangle 
\leftarrow 
N(\phi_k^{(n)}) \int d\vec x\,\frac{\tilde P(\vec x)}{N(\phi_k^{(n)})} 
B(\vec x) |\phi_k^{(n)}\rangle.
\label{eq:cpmcimpsamp}
\end{equation}
This iteration now forms the basis of the CPMC
algorithm. It is convenient to associate a weight $w_k^{(n)}$ with
each walker, which can be initialized to unity. One
step of the random walk is then as follows: For each walker
$|\phi_k^{(n)}\rangle$, {\bf ({\it i\/})} sample a $\vec x$ from
the probability density function $\tilde P(\vec x)/N(\phi_k^{(n)})$,
{\bf ({\it ii\/})} propagate the walker by $B(\vec x)$ to generate
a new walker, and {\bf ({\it iii\/})} compute a weight $w_k^{(n+1)}=
w_k^{(n)}N(\phi_k^{(n)})$ for the new walker.

Steps {\bf ({\it i\/})} and {\bf ({\it iii\/})} are sometimes
difficult to implement. To ease their implementation, we apply the HS
transformation of (\ref{eq:impcpmc}) and (\ref{eq:impcpmc_p}) to {\it
each\/} component of $\vec x$. This application is simple since both
$P(\vec x)$ and $B(\vec x)$ can be decomposed into a product of
independent factors corresponding to individual components
$x_\alpha$. Every step of the random walk then consists of successive
sub-steps in which the $x_\alpha$ are sampled one by one, each
according to {\bf ({\it i\/})} thru {\bf ({\it iii\/})}. As we discuss
in Section IV, such a decomposition is adequate to make the
Hubbard-model application straightforward, since the HS transformation
we use allows only two discrete values ($\pm 1$) for each $x_\alpha$,
and thus the easy tabulation of $\tilde P(x_\alpha)$. For more general
cases, however, it is often necessary to further approximate $\tilde
P(x_\alpha)$. The following procedure can be adopted: Under the
assumption of small $\Delta \tau$, the ratio of the overlap integrals
is manipulated into the form of an exponential whose exponent is
linear in $x_\alpha$; $\tilde P(x_\alpha)$ is then written as a
shifted Gaussian times a normalization constant. The basic idea of
this procedure is similar to that used in the DMC method.

To better see the effect of importance sampling, we observe that if
$|\psi_T\rangle=|\psi_0\rangle$, the normalization $\int \tilde P(\vec
x)d\vec x$ is a constant. Therefore the weights of walkers remain a
constant and the random walk has no fluctuation. Furthermore, we refer
again to the estimator for $E_0$. With importance sampling, the
denominator becomes the sum of weights $w$, while the numerator is
$\sum_{\phi} \langle\psi_T|H|\phi\rangle w/\langle\psi_T|\phi\rangle$,
where again $|\phi\rangle$ denotes walkers after equilibration. As
$|\psi_T\rangle$ approaches $|\psi_0\rangle$, all walkers contribute
equally to the estimator and the variance approaches zero. We
emphasize that different choices of importance functions only affect
the efficiency of the calculation.

\subsection{Constrained path approximation}

Despite the advantages over the standard AFQMC method in terms of
sampling efficiency, the random walk formulation still suffers from
the sign problem. Here we will illustrate the origin of the sign
problem in this framework and then introduce the constrained path
approximation to eliminate the exponential decay of the average
sign. We will see that while such a fixed-node-like approximation has
proved difficult to implement effectively in standard AFQMC method
\cite{Fahy}, it is extremely simple to impose under our random walk
formulation.

The sign problem occurs because of the fundamental symmetry existing
between the fermion ground-state $|\psi_0\rangle$ and its negative
$-|\psi_0\rangle$\cite{Fahy,inter}.  For any ensemble of Slater
determinants $\{|\phi\rangle\}$ which gives a Monte Carlo
representation of the ground-state wave function, this symmetry
implies that there exists another ensemble $\{-|\phi\rangle\}$ which
is also a correct representation.  In other words, the Slater
determinant space can be divided into two degenerate halves ($+$ and
$-$) whose bounding surface ${\cal N}$ is defined by
$\langle\psi_0|\phi\rangle=0$ and is in general {\it unknown\/}.

In some special cases, such as the particle-hole symmetric,
half-filled one-band Hubbard model, symmetry prohibits any crossing of
${\cal N}$ in the random walk. The calculation is then free of the
sign problem\cite{Hirsch}.  In more general cases, walkers {\it can\/}
cross ${\cal N}$ in their propagation by $e^{-\Delta\tau H}$. The sign
problem then invariably occurs. Once a random walker reaches ${\cal
N}$, it will make no further contribution to the representation of the
ground state:
\begin{equation}
\langle \psi_0|\phi\rangle = 0 \ \Rightarrow \  
\langle \psi_0|e^{-\tau H}|\phi\rangle 
= 0 \ {\rm for \ any} \ \tau.
\label{eq:node}
\end{equation}
Paths that result from such a walker have equal probability of being
in either half of the Slater determinant space. Computed analytically,
they would cancel, but without any knowledge of ${\cal N}$, they
continue to be sampled in the random walk and become Monte Carlo
noise. At sufficiently large $n$, the Monte Carlo
representation of the ground-state wave function consists of an {\it
equal\/} mixture of the $+$ and $-$ ensembles, regardless of where the
random walks originated. The Monte Carlo signal is therefore
lost. The decay of the signal-to-noise ratio, i.e. the decay of the average
sign of $\langle\psi_T|\phi\rangle$, occurs at an exponential rate 
with imaginary time.

In this regard, the fermion sign problem appears very similar in
either the DMC, AFQMC, or CPMC algorithms.  The difference between the
algorithms is that in the DMC algorithm minus signs appear when
particles interchange positions in configuration space while in the
CPMC and AFQMC algorithms the orbitals must interchange.  The orbitals
are an extended quantity and hence, at least for systems near a
mean-field solution, the fermion sign problem is reduced.

To eliminate the decay of the signal-to-noise ratio, we impose the
constrained-path approximation. It requires that each random walker at
each step have a positive overlap with the trial wave function
$|\psi_T\rangle$:
\begin{equation}
\langle \psi_T|\phi_k^{(n)}\rangle > 0.
\label{eq:constraint}
\end{equation}
This yields an approximate solution to the ground-state wave function,
$|\psi^c_0\rangle =\sum_\phi |\phi\rangle$, in which all Slater
determinants $|\phi\rangle$ satisfy (\ref{eq:constraint}). From
(\ref{eq:node}), it follows that this approximation becomes {\it
exact\/} for an exact trial wave function
$|\psi_T\rangle=|\psi_0\rangle$.

As a consequence of the constrained-path approximation, the
ground-state energy $E_0^c$, computed by the estimator discussed in
Section III A, is an upper bound to the true value $E_0$.  To see
this, we introduce an anti-symmetrization operator $A_\phi$ in the
Slater-determinant space that extends any wave function defined in
half the space by $\sum_\phi |\phi\rangle$ into the whole space by
$\sum_\phi |\phi\rangle - \sum_{-\phi} |-\phi\rangle $.  Since $A_\phi
|\psi_0^c\rangle$ is an eigenfunction of the modified Hamiltonian
$H^c=H+V^c$, where $V^c$ is $\infty$ at ${\cal N}$ and $0$ elsewhere,
we have $H^c(A_\phi |\psi_0^c\rangle)=H(A_\phi |\psi_0^c\rangle)
=E_0^\prime (A_\phi |\psi_0^c\rangle)$. Both $A_\phi |\psi_0^c\rangle$
and $A_\phi |\psi_0\rangle$ reside in the same Slater-determinant
space and both are antisymmetric functions. Thus $E_0^\prime \ge
E_0$.  On the other hand, we recall that 
\begin{equation}
E_0^c\equiv 
\frac {\langle \psi_T | H|\psi_0^c\rangle}{\langle \psi_T |\psi_0^c\rangle} 
=\frac{\langle \psi_T 
|H\,A_\phi|\psi_0^c\rangle}{\langle \psi_T |\,A_\phi|\psi_0^c\rangle}.
\end{equation}
Therefore $E_0^c=E_0^\prime$ and $E_0^c\ge E_0$.

To implement the constrained-path approximation in the random walk, we
redefine the importance function by (\ref{eq:O_T}):
\begin{equation}
O_T(\phi)\equiv {\rm max}\{ \langle \psi_T| \phi\rangle,
0\}. 
\label{eq:imp_cp}
\end{equation}
This prevents walkers from crossing the trial nodal surface ${\cal N}$
and entering the ``$-$'' half-space as defined by
$|\psi_T\rangle$. 
In the limit $\Delta\tau\rightarrow 0$, (\ref{eq:imp_cp}) ensures that
the walker distribution vanishes smoothly at ${\cal N}$ and the
constrained-path approximation is properly imposed. With a finite
$\Delta \tau$, however, $\tilde P(\vec x)$ has a discontinuity at ${\cal
N}$ and the distribution does not vanish. We have found this
effect to be very small for reasonably small imaginary-time steps
$\Delta\tau$ . Nonetheless, we correct for it by modifying
$\tilde P(\vec x)$ near ${\cal N}$ so that it approaches zero smoothly
at ${\cal N}$. As we discuss in Section IV, the procedure is analogous
to the mirror-correction\cite{FN,qmc_book} used in the DMC method.

\subsection{Computing expectation values}

After the random walk has equilibrated, the distribution of random
walkers represents the ground-state
wave function $|\psi_0^c\rangle$ under the constrained-path
approximation.  Various expectation values can then be computed from a
population of these walkers and their weights. For example, the
ground-state energy is
\begin{equation}
E_0^c =
\frac{\langle \psi_T | H | \psi_0^c \rangle}
     {\langle \psi_T  | \psi_0^c \rangle} =
\frac{\sum_k w_k \langle\psi_T|H|\phi_k\rangle/\langle\psi_T|\phi_k\rangle}
     {\sum_k w_k},
\label{eq:mixed}
\end{equation}
where terms in the numerator
$\langle\psi_T|H|\phi_k\rangle/\langle\psi_T|\phi_k\rangle$ are given
by combinations of elements of the Green's function as defined in
(\ref{eq:G}). 

An estimator similar to (\ref{eq:mixed}), namely $\langle\psi_T|{\cal
O} |\psi_0^c\rangle/\langle\psi_T|\psi_0^c\rangle$, is easily obtained
for any other operator ${\cal O}$. In the GFMC method, this type of
estimator is referred to as the {\it mixed\/} estimator. We recall 
that the true expectation value of ${\cal O}$ with respect to
$|\psi_0^c\rangle$ is
\begin{equation}
\langle {\cal O}\rangle^c =
\frac{\langle\psi_0^c|{\cal O}|\psi_0^c\rangle}
                           {\langle\psi_0^c|\psi_0^c\rangle}.
\label{eq:trueO}
\end{equation}
For the energy, the mixed estimator is equivalent to the true
estimator (\ref{eq:trueO}), but this is not true for any ${\cal O}$
that does not commute with $H$. In this case,
it is sometimes possible to
improve the mixed estimator by the following linear
extrapolation\cite{GFMC}:
\begin{equation}
\langle {\cal O}\rangle_{\rm extrap} \approx
2 \langle {\cal O}\rangle_{\rm mixed} -
\langle {\cal O}\rangle_{\rm var},
\label{eq:extrap}
\end{equation}
where the variational estimate $\langle {\cal O}\rangle_{\rm
var}=\langle\psi_T|{\cal O}|\psi_T\rangle/\langle\psi_T|\psi_T\rangle$.

Even a good trial wave function $|\psi_T\rangle$ with a good
variational energy can sometimes fail to give a reasonable estimate
for certain correlation functions. In such cases, (\ref{eq:extrap})
will not be effective.  It is then imperative to compute
(\ref{eq:trueO}). To do this, we devised a scheme called {\it
back-propagation} (BP), the essence of which comes from the forward
walking (FW) technique\cite{fwd} in the GFMC method:
\begin{equation} \langle{\cal O}\rangle_{\rm BP} = \lim_{\tau \rightarrow
 \infty}
\frac{\langle \psi_T \exp ( - \tau H_c)| {\cal O}| \psi_0^c \rangle}
     {\langle \psi_T \exp ( - \tau H_c)| \psi_0^c \rangle}.
\label{eq:bpest}
\end{equation}
A subtle distinction, however, exists between back-propagation and
forward walking.  In back-propagation, $\langle
\psi_T \exp ( - \tau H_c ) | = \langle \psi_T|
\exp ( -\tau H_c )$ is restricted to ``constrained'' paths, i.e., those
paths that do not violate the constraint in the {\it original forward
direction} $\exp ( - \Delta\tau H_c )|\psi_0^c \rangle$.  The standard
ground-state DMC method has no sense of ``time,'' because movements in
configuration space are exactly reversible. The CPMC method, however,
has a time direction: a set of determinants which does not violate the
constraint at any time when going from right to left may indeed
violate it any even number of times when evaluated from left to right.

Because of this sense of direction, expression (\ref{eq:bpest}) may
not yield (\ref{eq:trueO}).  However, since $|\psi_0^c\rangle $ is
itself approximate, this issue is not crucial.  What is crucial is
that $\langle{\cal O}\rangle_{\rm BP}$ remains exact for an exact trial
wave function.  To demonstrate that it does, we will use perturbation
theory and start by considering a Hamiltonian
\begin{equation}
H^\prime = H + \lambda {\cal O},
\end{equation}
where ${\cal O}$ is the operator whose expectation value we seek.  We
then apply the CPMC method to the new Hamiltonian $H^\prime$ with a
new constraint governed by $|\psi_T^\prime\rangle$, where
\begin{equation}
|\psi_T^\prime\rangle = |\psi_T\rangle + \lambda | \delta
\psi_T\rangle,
\end{equation}
and $|\delta \psi_T\rangle$ is orthogonal to $|\psi_T\rangle$, i.e.,
$\langle\psi_T^\prime|\psi_T\rangle = \langle\psi_T|\psi_T\rangle$,
which is the standard boundary condition of perturbation theory. In
the limit of small $\lambda$, the constraint becomes identical to that
of the original Hamiltonian $H$. To first order in $\lambda$, we thus
simply regain the previous expression (\ref{eq:bpest}) for
$\langle{\cal O}\rangle_{\rm BP}$. The term proportional to
$|\delta\psi_T\rangle$ does not contribute because it is orthogonal to the
true ground state.

Numerically we compared $\langle{\cal O}\rangle_{\rm BP}$ with
$\langle{\cal O}\rangle^c$ in several simple cases and observed
reasonable agreement. In principle we can compute $\langle{\cal
O}\rangle^c$ by creating a separate walk for the left-hand wave
function, which propagates from $\langle \psi_T|$ with an appropriate
sense of direction, and then matching it with populations in the
regular (right-hand) walk. In practice, however, it is difficult for
such a scheme to yield accurate results, due to a lack of proper
importance sampling\cite{O_c}.

Neither of the above measurement procedures is free from bias in the
long imaginary-time limit.  Since we are dealing with a branching
random walk, there is necessarily a bias that arises from finite
population sizes. However, this bias can be greatly reduced
by taking a relatively large population size (a few hundred
to a few thousand).  The convergence and amount of bias will depend
upon $|\psi_T\rangle$ and the low-energy excitations of the system. We
have found the back-propagation estimate of ground-state observables
to be both statistically and physically accurate in our Hubbard model
calculations.

It is relatively simple to implement the back-propagation scheme on
top of a regular CPMC calculation. We choose an iteration $n$ and
store the entire population $\{\,|\phi^{(n)}_k\rangle\,\}$.  As the
random walk proceeds, we keep track of the following two items for
each new walker: (1) the sampled auxiliary-field variables that led to
the new walker from its parent walker and (2) an integer that labels
the parent. After an additional $m$ iterations, we carry out the
back-propagation: For each walker $l$ in the $(n+m)^{\rm th}$
(current) population, we initiate a determinant $\langle \psi_T|$ and
act on it with the corresponding propagators, but taken in reverse
order. The $m$ successive propagators are constructed from the stored
items, with ${\rm exp}(-\Delta\tau K/2)$ inserted where necessary. The
resulting determinants $\langle \bar\phi^{(m)}_l|$ are combined with
its parent from iteration $n$ to compute $\langle{\cal O}\rangle_{\rm
BP}$, in a way similar to the mixed estimator (\ref{eq:mixed}). The
weights are given correctly by $w_l^{(n+m)}$, due to importance
sampling in the regular walk. Starting from another iteration $n'$,
this process can be repeated and the results accumulated.

\section{IMPLEMENTATION ISSUES: THE HUBBARD MODEL}

The one-band Hubbard model is a simple paradigm of a system of
interacting electrons. Its Hamiltonian is given by
\begin{equation}
  H = K+V =-t\sum_{\langle ij \rangle \sigma} (c_{i \sigma}^\dagger
    c_{j\sigma} + c_{j \sigma}^\dagger c_{i\sigma}) + U \sum_i n_{i
    \uparrow} n_{i \downarrow},
\label{eq:H}
\end{equation}
where $t$ is the overlap integral, $U>0$ is the on-site Coulomb repulsion,
$n_{i\sigma}=c_{i \sigma}^\dagger c_{i\sigma}$, and $\langle\;\rangle$
indicates near-neighbors. We will take $t=1$ and assume 
a two-dimensional square
lattice of size $N=L \times L$, with periodic boundary conditions.

The physics of this model is rich, containing magnetism, a
metal-insulator transition, and heavy Fermion behavior.  Originally
the model was proposed for ferromagnetism; today's interest focuses on
the extent to which it might exhibit superconductivity away from the
half-filled case of $N_{\uparrow}=N_{\downarrow}=N/2$.  It is in the
electron and hole-doped regions around half-filling that existing QMC
methods experience a debilitating sign problem that restricts the
simulations to small lattice sizes. In another paper, we will detail
our study of the physical properties of this model \cite{zhang96}.

To illustrate the CPMC algorithm in more detail, we now describe our
implementation of it for the Hubbard Hamiltonian.  For this and
related lattice models, we use the discrete version \cite{Hirsch} of
the Hubbard-Stratonovich transformation
\begin{equation}
  e^{-\Delta\tau U n_{i \uparrow} n_{i \downarrow}} =
   e^{-\Delta\tau U (n_{i\uparrow}+n_{i \downarrow})/2} \sum_{x_i=\pm 1}
     p(x_i)\, e^{\gamma x_i (n_{i \uparrow}-n_{i \downarrow})},
\label{eq:HSdiscrete}
\end{equation}
where $\cosh(\gamma) = {\rm exp}(\Delta\tau U/2)$ and the probability density
function $p(x_i)=1/2$ allows only  $x_i=\pm1$. Here we 
label components of the auxiliary field $\vec x$ by $i$, instead of by
$\alpha$, because of their one-to-one correspondence with lattice
sites.

\subsection{Specific issues}

Each Slater determinant $|\phi\rangle= 
|\phi^\uparrow\rangle|\phi^\downarrow\rangle$, and similarly $B(\vec
x)= B^\uparrow(\vec x) B^\downarrow(\vec x)$.  Any overlap integral
between two Slater determinants involves the product of overlaps of
individual spin determinants, e.g., $\langle \psi_T|
\phi\rangle=\prod_\sigma \langle \psi_T^\sigma| \phi^\sigma\rangle$.
Reflective of the interaction, the $\uparrow$ and $\downarrow$ spin
determinants share auxiliary fields. Aside from this connection, they
propagate independently.  With these details about electron spin taken
into consideration, all our previous discussions directly apply.

To reduce errors associated with the first-order Trotter approximation
(\ref{eq:trotter}), we use the second order symmetric form,
$\exp(-\Delta\tau H) \approx
\exp(-\Delta\tau K/2)\exp(-\Delta\tau V)\exp(-\Delta\tau K/2)$.
Thus,
\begin{equation}
  B ({\vec x}) = B_{K/2} B_V(\vec x) B_{K/2},
\label{eq:Bfactors}
\end{equation}
where $B_V(\vec x)$ is the auxiliary-field-dependent propagator from
the HS transformation. From (\ref{eq:HSdiscrete}), $B_V^\sigma(\vec
x)=\prod_i b_V^\sigma(x_i)$, where 
\begin{equation}
b_V^\sigma(x_i)= 
e^{-[\Delta\tau U/2+s(\sigma)\gamma x_i] c_{i\sigma}^\dagger
c_{i\sigma} },
\end{equation}
with $s(\uparrow)=1$ and $s(\downarrow)=-1$, and
$B_{K/2}^\sigma=e^{-\Delta\tau K^\sigma/2}$. The probability density
function is $P(\vec x)= \prod_i p(x_i)$.

For a walker $|\phi\rangle$, we now describe 
one sub-step of the random walk in which the $i^{\rm th}$ component
$x_i$ of the HS field is sampled and then $b_V(x_i)$ 
is applied to the walker.  If $b_V(x_i) |\phi\rangle$ is denoted by
$|\phi^\prime\rangle$, we have
\begin{equation}
  \tilde p(x_i) =
 \frac{O_T(\phi')}{O_T(\phi)}p(x_i).
\label{eq:pdiscrete}
\end{equation}
Thus the probability for picking one of the two possible values of
$x_i$ is $\tilde p(x_i)/[\tilde p(+1)+\tilde p(-1)]$. Once an $x_i$ is
chosen, $|\phi\rangle$ is propagated to obtain a new walker and the
new weight is $w\, [\tilde p(+1)+\tilde p(-1)]$.  Since ${\bf
b}_V^\sigma(x_i)$ modifies only one row of the matrix $\Phi^\sigma$,
it is straightforward and inexpensive to compute the resulting ratio
of determinants for each of the two possible values of $x_i$, provided
the inverse of the overlap matrix $(\Psi_T^{\rm T}\Phi^\sigma)^{-1}$
is known. As one moves to the next $i$, the overlap matrix can be
efficiently updated by procedures that are almost identical to ones
used in the AFQMC method\cite{BSS,LGbook}.

As we mentioned in Section III B, with a finite $\Delta\tau$, equation
(\ref{eq:imp_cp}) causes $\tilde p(x_i)$ to be discontinuous at ${\cal
N}$. To correct for this, we include a simple ``mirror correction''
procedure in the above sub-step: If walker $|\phi\rangle$ is close to
${\cal N}$, where we define ``close'' as when one application of
$b_V(x_i)$ (with either $x_i$) would lead to a determinant
$|\phi^\prime\rangle$ with a negative overlap with $|\psi_T\rangle$,
we modify its weight to $w/[1-\langle
\psi_T|\phi^\prime\rangle/O_T(\phi)]$.  After the new walker
$|\phi^\prime\rangle$ has been accepted, we again check whether it is
close to ${\cal N}$. This is done by applying $b_V(x_i)$ one
additional time with the chosen $x_i$. Similar to $w$ above, the
weight $w^\prime$ is modified if the overlap
$\langle\psi_T|b_V(x_i)|\phi^\prime\rangle$ is negative. We note that
there is essentially no computational cost in computing this overlap.

\subsection{Algorithm outline}

The basic steps of the algorithm are:
\begin{enumerate}
\item For each walker, specify its initial state
 to be some appropriate $\Psi_T$ and assign its weight $w$ and overlap
$O_T$ each a value of  unity.
\item If the weight of the walker is non-zero, propagate it via $B_{K/2}$: 
 \begin{enumerate} 
   \item Perform the matrix-matrix multiplications
    \begin{equation} 
      \Phi^\prime = {\bf B}_{K/2}\Phi 
    \end{equation} 
         and compute the new importance function 
    \begin{equation} 
      O_T^\prime = O_T(\phi^\prime).
    \end{equation} 
   \item If $O_T^\prime \ne 0$, update the walker, weight, and $O_T$:
    \begin{equation} 
    \Phi \leftarrow \Phi^\prime, \quad w \leftarrow w\, O_T^\prime/O_T,
       \quad O_T \leftarrow O_T^\prime .
    \end{equation} 
 \end{enumerate}
\item If the walker's weight is still non-zero, propagate it via $B_V(\vec x)
 =\prod_j b_V(x_j)$: 
 \begin{enumerate} 
   \item Compute the inverse of the overlap matrix 
      \begin{equation} 
         {\bf O}_{\rm inv} = (\Psi_T^{\rm T} \Phi)^{-1}.  
      \end{equation} 
   \item For {\it each\/} auxiliary field $x_i$ ($i=1,N$)
      \begin{enumerate} 
         \item compute $\tilde p(x_i)$ in which the ratio with $O_T$ 
               can be expressed in terms of $x_i$ and the Green's function
               $G_{ii}$ as defined in (\ref{eq:G});
         \item sample $x_i$ and update the weight according to 
               the discussion in Section IV A;
         \item if the weight of the walker is still not zero, propagate
               the walker by ${\bf b}_V(x_i)$ and 
               then update $O_T$ and ${\bf O}_{\rm inv}$;
         \item apply the mirror correction in steps i or iii if necessary.
      \end{enumerate} 
 \end{enumerate}
\item Repeat step 2.
\item Include an overall normalization factor in walker weight:
  $w=w\, e^{\Delta\tau E_T}$, where $E_T$ is an estimate of $E_0$.
\item Repeat 2 thru 5, which form one step of the random walk, 
for all walkers in the population.
\item If the population of walkers has achieved a steady-state
 distribution, periodically make estimates of physical quantities.
\item Periodically adjust the population of walkers.
\item Periodically re-orthonormalize the columns of all the $\Phi$
 with non-zero weights.
\item Cycle the process until an adequate number of measurements are
 collected.
\item Compute final averages, estimate their statistical error, and stop.
\end{enumerate}

We omitted the spin index $\sigma$, but identical operations for
both electron spin determinants are implied whenever a matrix
manipulation is described. 
In presenting the above steps, we focused on illustrating the
algorithm; the outline does not represent the most efficient 
implementation. 

We also ignored back-propagation. To implement it, an additional
procedure is required in each random walk step in the measurement
phase. As discussed at the end of Section III C, this procedure takes
{\it one\/} of the following three possibilities: ({\it i\/}) storing
all walkers (step 6), ({\it ii\/}) storing ancestry links (step 3) and
auxiliary fields (step 3.b.iii) for each walker, and copying them if
necessary (step 8), or ({\it iii\/}) back-propagating from $\langle
\psi_T|$, computing $\langle{\cal O}\rangle_{\rm BP}$, and
accumulating results (step 6). The length $m$ of the back-propagation is
pre-set as are the number of iterations in each measurement block,  the
frequency of measurements, etc.

Additionally, we introduced several previously undiscussed steps and
procedures. We now discuss these.

\subsection{Additional general algorithmic issues}

In this section, we discuss the use of $E_T$, the growth estimator,
population control, and re-orthogonalization. Variants of the first
three are present in all GFMC calculations, while the last one is
adopted from the AFQMC method. We merely highlight these issues in the
context of the new algorithm and point to references with more
extensive discussions. While we will refer to the outline for the
Hubbard implementation in Section IV B, the issues are general.

As shown in step 5, a constant $e^{\Delta\tau E_T}$ multiplies the
propagator $e^{-\Delta\tau H}$. If $E_T=E_0^c$, the steady-state
eigenvalue of equation (\ref{eq:process}) is unity. In other words,
after equilibration the total weight of walkers remains a constant on
the average. An inaccurate input of $E_T$ is only a minor concern,
since the value of $E_T$ simply changes the total weight
systematically by a constant factor. A statistical estimate of this
factor is easily obtained from the random walk by observing the total
weights through a number of iterations\cite{GFMC}. On the average
these totals scale as ${\rm exp}[-\Delta\tau (E_0^c-E_T)]$ and hence
provide an estimate of $E_0^c$. This procedure for estimating the
ground-state energy is often referred to as the {\it growth
estimator\/}.
 
In the random walk, one walker will eventually dominate all others and
the random walk will spend most of its time sampling walkers which
contribute little. To avoid such a loss of efficiency, a population
control procedure\cite{qmc_book} is needed (step 8).  First, a
branching (or birth/death) scheme is applied, in which walkers with
large weights are replicated and ones with small weights are
eliminated by some probability. There exist various ways to do
this\cite{GFMC,qmc_book,Hetherington}, with the guideline being that
the process should not affect the distribution
statistically. Branching allows the total number of walkers to
fluctuate and possibly become too large or too small. Thus as a second
step, the population size is adjusted, if necessary, by rescaling the
weights with an overall factor. Re-adjusting the population size
introduces a bias and should only be done infrequently. If this is not
possible due to a poor $|\psi_T\rangle$ or poor importance sampling,
several calculations must be done with different (average) population
sizes in order to extrapolate to the infinite population limit.

The overall structure of CPMC resembles a typical GFMC
calculation\cite{GFMC}. After equilibration, we introduce an
intermediate phase in which the growth estimator is computed. The
length of this phase is a parameter and the outcome is used to adjust
$E_T$. The new value of $E_T$ is then used in the next phase, which is
divided into independent blocks.  Because of the correlated nature of
successive steps, measurements are made only at suitable intervals.
At an iteration $n$ when measurements are taken, the average of a
quantity ${\cal O}$ is ${\cal O}^{(n)}=u^{(n)}/d^{(n)}$. In each
block, we do not accumulate ${\cal O}^{(n)}$. Instead we accumulate
$u^{(n)}$ and $d^{(n)}$ separately and compute a final average at the
end of the block. The final result and statistical error estimate
(step 11) are obtained from these bin averages.

Step 9 is prompted by the fact that the repeated multiplication of
${\bf B}(\vec x)$ leads to a numerical instability: relatively quickly
numerical error grows to the point where $|\phi_k^{(n)}\rangle$
represents an unfaithful propagation of $|\phi_k^{(0)}\rangle$.  This
instability is well-known in the AFQMC method and is
controlled\cite{LGbook,white} by a numerical stabilization technique
that requires the periodic re-orthonormalization of the single
particle orbitals in $|\phi_k^{(n)}\rangle$. In our calculation, we
use the {\it modified} Gram-Schmidt procedure\cite{LGbook,white} which
for each walker $|\phi\rangle$ factors the matrix $\Phi$ as $\Phi =
{\bf QR}$, where ${\bf Q}$ is a matrix whose columns are a set of
orthonormal vectors and ${\bf R}$ is a triangular matrix. After this
factorization, $\Phi$ is replaced by ${\bf Q}$ and the corresponding
overlap $O_T$ is replaced by $O_T/\det({\bf R})$.

\section{RESULTS}

   We have presented a rather detailed discussion of the constrained
path Monte Carlo method to make the CPMC algorithm as transparent as
possible.  In this section we present a variety of results from
simulations of the Hubbard model that are chosen primarily for their
importance in illustrating algorithmic issues.  These results are for
two-chain and two-dimensional lattices. First, we give results for the
ground-state energy as a function of system size $N$, filling fraction
$\langle n \rangle = (N_\uparrow + N_\downarrow)/N$, and trial wave
function.  Later, we examine results for other physical quantities
such as pairing correlation functions.

\subsection{Ground-State Energy}

   In Figure 1 we demonstrate the convergence and stability of the
CPMC method.  The energy (top) and overlap integral (bottom) are
plotted as a function of imaginary time $\tau\equiv n\Delta\tau$. On
the left, we demonstrate the initial convergence from the trial wave
function to the ground state, and on the right, the asymptotic
stability of the algorithm.  This figure is constructed for an $8
\times 8$ lattice
with $\langle n \rangle = 0.875$, and $U = 4$. This
parameterization generates a fairly
difficult system for the AFQMC method because of the sign problem. 
$|\Psi_T\rangle$ was an unrestricted Hartree-Fock wave function
obtained with $U=0.4$ ({\it not\/} $4$) and yielded a variational
energy of $-53.05$. The average number of walkers was
$600$.  The behavior
shown is quite typical of CPMC calculations. 

   In the relaxation phase ($\tau \le 10$), a trial energy of
$E_T = -58.0$ was used.  This value is significantly higher than the
true ground-state energy, so the overlap integral $\langle \psi_T | \psi^{(n)}
\rangle $, and consequently the population size, grows quite
rapidly. During this phase, population control is applied frequently
(every $5$ steps).  The rapid fluctuations in value of the overlap
integral shown in the bottom left are due to re-adjustments of the
total population size in each of these applications. For $10<\tau\le
20$ (not shown), we used the same $E_T$ and applied population control
with the same frequency, but computed the ground-state energy with the
growth estimator. This value was then used for the trial energy $E_T$
during the measurement phase ($\tau > 20$). In this phase, even though
population control was applied every $10$ steps and branching occurs,
the total population size remained within the pre-set lower and upper
bounds of $300$ and $1200$. Indeed, all vertical displacements in the
overlap integral in the lower right of the figure correspond to the
beginning of a block where we reset the population size to within
$10\%$ of the expected average of 600.

Clearly, the values of both the overlap integral and the energy are
completely stable as a function of $n$. In contrast, the standard
AFQMC method has a bad sign problem.  Indeed, for a $4\times 4$
lattice at the same filling fraction and the same $U$, the average
sign decays exponentially by roughly $5$ {\it orders of magnitude\/}
as the imaginary time increases from $0$ to $20$ \cite{sign1}.

The ground-state energy $E_0^c=-65.135\pm 0.008$. (Extrapolation to
$\Delta\tau=0$ will slightly reduce the average value.) The best AFQMC
result available for this system is $-65.02\pm 0.06$\cite{Imada},
which is slightly higher than our upper bound.  The $\tau$-dependence
of the energy after convergence is shown in the upper right of the
figure. The inset demonstrates the short-term fluctuations in the
mixed estimate of the energy; these fluctuations are also shown as the
solid line in the main figure.  Constructing local ``binned'' averages
of the energy from blocks of length $\tau=10$ yields the energy
estimates shown as dotted lines in the inset. These binned averages
are nearly statistically independent, and their root mean squared
deviations divided by the square-root of the number of bins yielded
the quoted statistical error.

   The computational requirements of the CPMC method are fairly modest
when compared, for example, with the AFQMC method.  Efficiency,
however, can be dramatically affected by implementation issues.  Two
important issues are the accuracy of the trial wave function and of
the imaginary-time propagator.  To obtain the maximum efficiency, it
is essential to use a propagator accurate to second order in the
break-up between kinetic and potential terms.  For example, figure 2
shows the energies obtained for first-order and second-order Trotter
approximations for the propagators for a $4\times 4$ lattice with
$N_\uparrow=N_\downarrow=6$ spins and $U=4$.  The first-order
propagator introduces significant systematic effects even for quite
small time steps, however, the second order propagator permits
significantly larger values of $
\Delta\tau$ to be used. We note that we also included a mirror
correction to the 
propagation to make certain that there are no corrections of order less
than ${(\Delta\tau)^2}$.  In this case, however, the differences between
using second-order propagators with and without this correction are
comparable to the error bars in the figure.



   Of course, while stability and efficiency are necessary for a
useful simulation, accuracy is the principal concern. In Tables I and
II, we compare our results for the Hubbard model to exact results for
small systems and to other methods for large systems. Since the CPMC
method depends upon approximate knowledge of the ground state of the
system, we also included results for different trial wave functions.
To date, we have used only free-electron and unrestricted Hartree-Fock
(uHF) trial wave functions.

   In Table I, we see that the accuracy of the CPMC ground-state
energy is always better than 5\%, often much better, even when the
trial wave function is very poor.  The worst case is $3\times3$ with
$4$ $\uparrow$ and $4$ $\downarrow$ spins, which is an open-shell case
that corresponds to a very difficult filling fraction for the AFQMC
method, and has a large $U$ of $8$, which makes single-determinant
trial wave functions rather poor approximations. The Hartree-Fock wave
function used as $|\psi_T\rangle$ had a very poor energy, $-0.0025$,
compared to the exact energy of $-0.809$.  The CPMC method however was
still able to obtain an energy of $-0.766(2)$.  Clearly a more
accurate approximation to the ground state wave function would yield
an even better result.

   In Table II we compare our results with available data from other
numerical approaches, including stochastic diagonalization (SD)
\cite{deRaedt}, AFQMC, and density-matrix renormalization group (DMRG)
\cite{DMRG} methods.  The SD method uses Monte Carlo methods to 
attempt the construction of an efficient basis for approximating the
ground state wave function of the system.  Since an explicit basis is
used, no sign problem occurs; however, an exponential growth in
computing time occurs reflecting the increased effort in selecting
members of the basis as system size increases.  In contrast, the AFQMC
method is in principle exact, but suffers from exponential growth in
computing time as system size increases because of the sign problem.
Finally, the DMRG method is a variational method that is very
effective for one-dimensional and quasi-one-dimensional models.

In Table II we focus filling fractions $\langle n\rangle $ around or
greater than $80\%$ as they are more interesting and also more
difficult for QMC calculations. For the closed-shell $4\times 4$
system, the results from the CPMC, SD, and AFQMC methods agree well.
As we increase the lattice size, the SD results are comparatively
poorer, presumably because of an insufficient number of states.  The
results from the AFQMC and CPMC methods continue to agree well up to
fairly large lattice sizes. The worst case is an $8\times 8$ lattice
with $25$ $\uparrow$ and $25$ $\downarrow$ spins, where the CPMC
result lies approximately 0.4\% $\pm$ 0.2\% above the AFQMC
result. For still larger lattice sizes, the sign problem limits the
use of the AFQMC method to only closed-shell systems, for which the
sign problem is much reduced. Even in these cases, we see that the
error estimates in AFQMC are much larger than the corresponding
statistical errors from the CPMC method. In the $12\times 12$ case,
the difference is roughly a factor of $30$, i.e., about a factor of
$900$ more in CPU time. (Our calculations typically took tens of hours
on an IBM RS6000 590.) Furthermore, the CPMC result is actually lower
in energy in this case than the AFQMC results, but size of the error
bars in AFQMC are similar to the difference between the two
results. The CPMC result on a $16\times 16$ lattice, a size far beyond
the reach of the AFQMC method due to the sign problem, was obtained
from a simulation comparable to that for the $12\times 12$ system in
terms of the numbers of iterations and walkers.  For the two-chain
system, we obtained excellent agreement with the DMRG results of Noack
{\it et al\/}. Here the energy agreed to within less than 0.1\%.

\subsection{Correlation Functions}

    To effectively study ground-state properties, we need to
accurately calculate pairing correlation functions, momentum
distributions, and other ground-state expectation values. In previous
sections, we discussed various ways of estimating a ground-state
expectation value, in particular the back-propagation scheme. Here we
will mostly benchmark results and further discuss related algorithmic
behavior. We remark that for any simulation method correlation
functions are in general much more difficult to compute than the
energy. Thus there is limited data available from other methods with
which we can benchmark our correlation functions. This means
self-consistency checks (e.g., comparison of results obtained with
different choices of $|\psi_T\rangle$) are crucial.

In Table III, we show results on the simple closed-shell $4\times 4$
system of $5\uparrow\,5\downarrow$ electrons at $U=4$ and with
$\Delta\tau=0.05$.  The one-body density matrix is the expectation
value of the Green's function elements: $\rho({\bf l})=\langle c_{\bf
0}^\dagger c_{\bf l}\rangle$, where ${\bf l}=(l_x,l_y)$. The spin
density structure factor is
\begin{equation}
S(k_x,k_y) = S({\bf k}) = 1/N\sum_{\bf l} {\rm exp}(i{\bf k}\cdot {\bf l})
\langle {\bf s}_{\bf 0} {\rm s}_{\bf l}\rangle,
\label{eq:struct}
\end{equation}
where ${\bf s}_{\bf l}=n_{{\bf l}\uparrow}-n_{{\bf l}\downarrow}$ is the
spin at site ${\bf l}$. The charge density structure factor $S_d$ is similar to
(\ref{eq:struct}), with spin replaced by density, i.e., with the $-$ sign
in ${\bf s}_{\bf l}$ replaced by a $+$ sign. The electron pairing
correlation is defined as
\begin{equation}
D_\alpha (l_x,l_y) = D({\bf l}) = \langle \Delta^\dagger_\alpha({\bf l})
\Delta_\alpha({\bf 0})\rangle,
\label{eq:pairing_def}
\end{equation}
where $\alpha$ indicates the nature of pairing. The on-site $s$-wave
pairing function has $\Delta_{1s}({\bf l}) = c_{{\bf l}\uparrow} c_{{\bf
l}\downarrow}$, while in this case for $d$-wave we used $\Delta_{2d} ({\bf l})
=c_{{\bf l}\uparrow}\sum_{\bf \delta} f({\bf \delta}) c_{{\bf l}+{\bf
\delta}\,\downarrow}$, where ${\bf \delta}$ is $(\pm 1,0)$ and $(0,\pm
1)$. For ${\bf \delta}$ along the $x$-axis, $ f({\bf \delta})$ is $1$;
otherwise it is $-1$. We average over different ${\bf 0}$ sites to
improve statistics.

In Table~III we compare CPMC results from the back-propagation
estimate with exact results. (The length of back-propagation was
$\tau=6$.) We see that the CPMC result is essentially exact for this
system.  The variational results suggest that the free-electron
$|\psi_T\rangle$ is not a very good trial wave function. Nonetheless
the constrained-path error seems to be negligibly small. In fact, very
limited branching occurs in the calculations, which indicates
effective importance sampling with $|\psi_T\rangle$.  Furthermore, as
$U$ is increased to $8$, the variational results become worse, but the
CPMC results obtained with the same $|\psi_T\rangle$ remain accurate
\cite{PRL}. Also in this table are estimates by the mixed and
extrapolation schemes discussed in III.C. While these schemes often
improve the variational results considerably, their systematic errors
are significant, particularly when compared to the high level of
statistical accuracy that can be achieved with the CPMC method.

In Table IV, we show expectation values for a system of
$7\uparrow\,7\downarrow$ electrons. This open-shell case has the worst
sign problem for a $4\times 4$ system. We show results from CPMC
simulations with two different trial wave functions. Both are
unrestricted Hartree-Fock wave functions, but $|\psi_{T1}\rangle$ was
obtained with a $U$ of $0.1$, while $|\psi_{T2}\rangle$ with
$U=4$. The calculation with $|\psi_{T1}\rangle$ has much less
fluctuation, even though $|\psi_{T2}\rangle$ has a lower variational
energy. In fact, we found this trend to be rather general:
free-electron-like wave functions tend to be better importance
functions than unrestricted Hartree-Fock wave functions. We see that
the two trial wave functions yield very different variational
estimates, but their CPMC results are consistent and in reasonable
agreement with exact results. For example, in the free-electron-like
$|\psi_{T1}\rangle$, the momentum distribution is a step function, and
the ${\bf k}=(1,0)$ state is completely occupied ($n_k=1$), but with
this trial wave function the CPMC method still gives the correct
occupation of $0.92(1)$.

   In Figure 3 we show the $d$-wave pairing correlation function
$D_{2d}({\bf l})$ and static magnetic structure factor $S({\bf k})$
obtained with two different trial wave functions for a $6\times 6$
system. The expectation values obtained directly from these two
different trial wave functions are shown in thick lines, and the
corresponding back-propagation estimates $\langle{\cal O}\rangle_{\rm
BP}$ are shown with thin lines.  While the two CPMC estimates do not
agree exactly, they do demonstrate a much closer correspondence with
each other than those obtained with the original wave functions. The
case shown is comparatively easy because of the small size of the
lattice and the relatively low value of electron filling. However, the
overall trend is rather general.

\section{SUMMARY AND DISCUSSION}

We described in detail the background, formalism, and implementation
of the new constrained path Monte Carlo algorithm. The CPMC method is
a general quantum Monte Carlo algorithm for computing fermion
ground-state properties. It introduces several new concepts, including
importance-sampled random walks in a Slater-determinant space and the
constrained path approximation within this framework. The algorithm
combines advantages of the existing Green's function Monte Carlo and
auxiliary-field quantum Monte Carlo methods, is free of any
signal-to-noise ratio decay, and scales algebraically with system
size. Together with data in Ref.~\cite{PRL}, we demonstrated that the
method produces very accurate results for the Hubbard model, even with
very simple choices of the trial wave function $|\psi_T\rangle$.

Compared to the GFMC method, the current algorithm allows the random
walk to take place in a basis other than that of configurations or
occupation numbers. In this sense, the CPMC algorithm is a
generalization of the GFMC algorithm. The CPMC method expresses the
ground-state wave function (stochastically) in the form of
(\ref{eq:wf}), rather than $\psi_0(R)=\sum_k
\delta(R-R_k)$ as in the GFMC method. This is advantageous as it makes 
feasible the use of various techniques developed for one-electron
calculations for atoms and solids.  In addition, it makes our
back-propagation scheme efficient and effective. Thus expectation
values can be computed via (\ref{eq:G}), while in the GFMC method the
analogous forward walking technique has often been difficult and
computations of some correlation functions have almost been
impossible. If applications of the CPMC method to continuum systems
are successful, the ability to compute expectation values such as
forces will be very valuable. Such applications are under study.

There is an obvious resemblance between the constrained path (CP)
approximation in the CPMC method and the fixed-node (FN) approximation
in the GFMC method. Both result in solutions to the Schr\"odinger
equation that are consistent with some artificial boundary
conditions. An important difference, however, is also evident. The FN
approximation is in configuration space and requires the solution to
have a pre-defined node: $\psi_0^{\rm FN}(R)=0$ where
$\psi_T(R)=0$. The CPMC method is in a Slater determinant space. The
CP approximation on each individual Slater determinant $|\phi\rangle$,
{\it translated\/} into configuration space $|R\rangle$, is non-local
and requires $\int\psi_T(R)\phi(R)dR >0$. Thus the node, as well as
the amplitude $\phi(R)$, is allowed to vary.  The systematic error in
the CPMC method arises because the solution it yields, in the form of
(\ref{eq:grdwf}), has the artificial constraint
$\chi_{\psi_0}(\phi)>0$.

In the CPMC method, each Slater determinant $|\phi\rangle$
analytically defines a continuous function $\phi(R)$ in contrast to
walkers in the GFMC method which are delta-functions. It is thus
easier to impose symmetries. One example is the trivial case of a
non-interacting system: the CPMC method naturally yields the correct
result, while standard GFMC still requires knowledge of the node.
Another is the one-band Hubbard model at half-filling, where the CPMC
method retains the exact nature of the AFQMC method, while the GFMC
method retains the sign problem\cite{half_fill_imp}.

Recently, ten Haaf and van Leeuwen\cite{FNlatt_new} presented data on
the Hubbard model from standard GFMC simulations with the fixed-node
approximation, which they and collaborators had earlier
generalized\cite{FNlatt} to treat lattice fermion systems. For the
$4\times 4$ system in Table II ($5\uparrow\,5\downarrow$ $U=4$), the
CPMC result for the energy per site is $E/N=-1.2239(3)$ (exact value
\cite{diag}: $-1.2238$). With an {\it identical\/} trial wave
function, the fixed-node calculations of ten Haaf and van Leeuwen
yielded $-1.2186(4)$.  Incorporating a Gutzwiller factor only slightly
improved their FN result to $-1.2201(4)$. Unfortunately, the rest of
their results are all FN energies computed for half-filled systems,
for which the CPMC energies would be {\it exact\/}.  Further
comparisons away from half-filling would allow a more systematic
understanding of the relative strengths of the CPMC method.

It is worth noting that the CPMC algorithm provides a stochastic
method closely linked with more traditional quantum chemistry
approaches such the configuration interaction (CI) method. Similar to
the CI method, the CPMC method produces a collection of determinants
whose sum represents the ground-state wave function. The determinants,
however, do not have to be orthogonal to each other. Furthermore, they
are generated efficiently and {\it systematically\/} by a Monte Carlo
process that is {\it guided} by importance sampling. The drawback of
the CPMC method is of course its variational nature due to the CP
approximation.

We are currently investigating several schemes for further improving
the algorithm. These include a method analogous to released-node
technique \cite{FN} in the GFMC method. For the energy, this seems
straightforward.  For other expectation values, it involves evaluating
$\langle {\cal O}\rangle_{\rm BP}(\tau_1,\tau_2)$ defined as:
\begin{equation}
\langle{\cal O}\rangle_{\rm BP} (\tau_1,\tau_2) 
= \langle \psi_T\,\exp [ - \tau_2 H^c] \exp [ -\tau_1 H] |
{\cal O}| \exp [ -\tau_1 H] | \psi_0 \rangle.
\end{equation}
In this expression $H$ is the original Hamiltonian without the
constraint, and $H^c$ indicates the Hamiltonian in the presence of the
constraint.  Hence, for a period of twice $\tau_1$, we evolve the
system without constraint and for another $\tau_2$ we include the
constraint. In the limit of zero $\tau_1$ we obtain the approximation
used to date, while finite $\tau_1$ improves the estimate, i.e., makes
it exact, at the cost of increasing statistical error.  The
bookkeeping in such a calculation could be arranged to calculate
directly the difference between the current $\langle{\cal
O}\rangle_{\rm BP}$ and the transient estimation, and hence to provide
a stringent test on the accuracy of a given calculation.

Other possibilities for improving estimates of expectation values
include optimization techniques for improving the trial wave function.
As mentioned earlier, the algorithm as described can be used with a
multi-determinant $|\psi_T\rangle$, with the computational cost
increasing only linearly with the number of determinants. Thus it is
desirable to have good trial wave functions in the form of linear
combinations of Slater determinants. In addition, wave functions in
this form that can be tuned systematically to yield different
properties would be highly useful, since self-consistency checks with
the CPMC method can then be carried out simply by changing parameters
in $|\psi_T\rangle$. Yet other algorithmic topics include the
development of interacting-walker \cite{inter} and mirror potential
\cite{mirrorpot} analogs.

\acknowledgements

We thank A.~Moreo, R.~M.~Noack, and R.~T.~Scalettar for providing
unpublished data, M.~H.~Kalos, D.~J.~Scalapino, and C.~J.~Umrigar
for stimulating discussions, and M.~Guerrero, M.~M.~Steiner and J.~W.~Wilkins
for helpful comments on the manuscript.  This work was supported
in part by the Applied Mathematics Program of
the Department of Energy. Calculations were performed at the Cornell
Theory Center on the SP2 computer. SZ also acknowledges support by
DOE-BES, Division of Material Science (DE-FG02-88ER45347).

%

\newpage

\begin{center}
\begin{table}
\tabcolsep=0.1in
\caption{Hubbard model ground-state energies per site from CPMC
simulations compared
with exact results. The first column under ``system''
is the lattice size; the second, the numbers of electrons with $\uparrow$
and $\downarrow$ spins. $U$ is the on-site Coulomb repulsion.
The trial wave function $|\psi_T\rangle$ used in CPMC is either a
free-electron (free) or an unrestricted Hartree-Fock (uHF) wave
function. $E_{\rm var}$ indicates the corresponding variational energy
from this wave function.  Statistical errors are in the last digit and
are shown in parentheses. Exact results for the $4\times 4$ systems
are taken from \protect\cite{diag,diag2,deRaedt}.}
\begin{tabular}{ccc r@{}l r@{}l r@{}l} 
system &
$U$ & $|\psi_T\rangle$ & 
\multicolumn{2}{c}{$E_{\rm var}/N$} & 
\multicolumn{2}{c}{$E_{\rm CPMC}/N$}  & 
\multicolumn{2}{c}{$E_{\rm exact}/N$} \\ \hline
2$\times$2 \ \ 2$\uparrow$ 1$\downarrow$&4&uHF&-1.&5327&-1.&6038(6)&-1.&6046 \\
2$\times$3 \ \ 2$\uparrow$ 2$\downarrow$&4&free&-1.&267&-1.&3828(9)&-1.&4009 \\
2$\times$3 \ \ 2$\uparrow$ 2$\downarrow$&8&free&-0.&889&-1.&221(2)&-1.&244 \\
2$\times$4 \ \ 2$\uparrow$ 2$\downarrow$&4&uHF&-1.&333& -1.&3678(5)&-1.&374 \\
2$\times$4 \ \ 3$\uparrow$ 3$\downarrow$&4&uHF&-1.&438&-1.&5693(5) &-1.&569 \\
1$\times$8 \ \ 3$\uparrow$ 3$\downarrow$&4&free&-0.&645& -0.&8329(7)&-0.&834 \\
3$\times$3 \ \ 4$\uparrow$ 4$\downarrow$&8&uHF&-0.&0025& -0.&766(2)& -0.&809 \\
4$\times$4 \ \ 2$\uparrow$ 2$\downarrow$&4&uHF&-2.&8225&-2.&8813(3)&-2.&8825 \\
4$\times$4 \ \ 4$\uparrow$ 4$\downarrow$&4&uHF&-1.&025&-1.&095(1) &-1.&096  \\
4$\times$4 \ \ 5$\uparrow$ 5$\downarrow$&8&free&-0.&7188&-1.&0925(7)&-1.&0944\\
4$\times$4 \ \ 6$\uparrow$ 6$\downarrow$&4&uHF&-1.&3117&-1.&4763(5)&-1.&478  \\
4$\times$4 \ \ 7$\uparrow$ 7$\downarrow$&4&uHF&-0.&8669&-0.&9831(6)&-0.&9838\\
4$\times$4 \ \ 7$\uparrow$ 7$\downarrow$&12&uHF&-0.&474&-0.&606(5)&-0.&628\\
\end{tabular}
\end{table}

\begin{table}
\caption{Hubbard model ground-state energies from CPMC
simulations compared
with available results from other approaches. The first
two columns follow the same convention as the corresponding ones in
Table I. The interaction strength $U$ is $4$. The stochastic
diagonalization (SD) results are from \protect\cite{deRaedt}; the
density-matrix renormalization group (DMRG) results on two-chains are
from \protect\cite{noackpri}. The statistical errors are in the last one
or two digits, as indicated.}
\begin{tabular}{c@{\ \ \ }c c r@{}l r@{}l r@{}l} 
\multicolumn{2}{c}{system} & $|\psi_T\rangle$ &
\multicolumn{2}{c}{$E_{\rm CPMC}$}  & 
\multicolumn{2}{c}{$E_{\rm SD}$} & 
\multicolumn{2}{c}{$E_{\rm AFQMC}$} \\ \hline
4$\times$4 & 5$\uparrow$ 5$\downarrow$ & free& -19.&582(5) & -19.&58 & -19.&58(1) \\
6$\times$6 & 13$\uparrow$ 13$\downarrow$ & free & -42.&34(2) & -40.&77 &     -42.&32(7)     \\
6$\times$6 & 14$\uparrow$ 14$\downarrow$ & uHF & -40.&17(2) &  &     & -40.&44(22)     \\
8$\times$8 & 25$\uparrow$ 25$\downarrow$ & free & -72.&48(2) & -67.&00 &     -72.&80(6)     \\
8$\times$8 & 27$\uparrow$ 27$\downarrow$ & uHF & -67.&46(4) &  &  & -67.&55(19)     \\
10$\times$10 & 41$\uparrow$ 41$\downarrow$ & free &  -109.&55(3) &  &   &     -109.&7(6)     \\
12$\times$12 & 61$\uparrow$ 61$\downarrow$ & free & -153.&43(5) &  &   &     -151.&4(1.4)     \\
16$\times$16 & 101$\uparrow$ 101$\downarrow$ & free &  -286.&55(8) &  &   & &\\
\hline
\multicolumn{2}{c}{system} & $|\psi_T\rangle$ &
\multicolumn{2}{c}{$E_{\rm CPMC}$}  & 
\multicolumn{2}{c}{$E_{\rm DMRG}$} & 
\\ \hline
2$\times$8 & 7$\uparrow$ 7$\downarrow$  & free &  -13.&067(4) &  -13.&0664(2) &    & \\
2$\times$16 & 14$\uparrow$ 14$\downarrow$ & free & -26.&87(2) &  -26.&867(3) &    & \\
\end{tabular}
\end{table}

\begin{table}
\tabcolsep=0.1in
\caption{Comparisons of the computed expectation values and 
correlation functions by different estimators for a $4\times 4$
lattice with
$5\uparrow\,5\downarrow$ electrons and $U=4$. The trial wave function
is the free-electron wave function. The corresponding variational
values from it are shown in the first row. The next row contains mixed
estimates from CPMC simulations, while ``extrp'' shows the
values extrapolated from 
the first two rows via (\protect\ref{eq:extrap}). The row labeled BP
gives the CPMC result via the back-propagation scheme.  In the
last row, $D_{1s}$ is from AFQMC simulations\protect\cite{deRaedt}, while the
others are exact diagonalization results, with the first four from
\protect\cite{diag} and the last from \protect\cite{Moreo}. $E_k$
indicates the kinetic energy, $\rho(l_x,l_y)$ the one-body density
matrix, $S$ and $S_d$ the spin and charge density structure factors, and
$D_{1s}$ and $D_{2d}$ the $s$- (on-site) and $d$-wave pairing
correlations respectively. The Monte Carlo errors are shown in
parentheses.}
\begin{tabular}{c r@{}l r@{}l r@{}l r@{}l r@{}l r@{}l}
 & 
\multicolumn{2}{c}{$E_k$} & 
\multicolumn{2}{c}{$\rho(2,1)$} & 
\multicolumn{2}{c}{$S(\pi,\pi)$} &
\multicolumn{2}{c}{$S_d(\pi,\pi)$} &
\multicolumn{2}{c}{$D_{1s}(2,1)$} & 
\multicolumn{2}{c}{$D_{2d}(2,1)$} \\ \hline
variational&-24.&0 &-0.&0625 & 0.&625 & 0.&625 & 0.&003906 & 0.&03125\\
mixed &-24.&0(0)& -0.&0625(0)& 0.&6938(4)& 0.&5572(1)& 0.&000684(3) 
 & 0.&03095(2)\\
extrap & -24.&0(0) & -0.&0625(0) & 0.&763(1) & 0.&4894(2) & -0.&002538(6) 
 & 0.&03065(4)\\
BP & -22.&55(2) & -0.&0563(3) & 0.&729(1) & 0.&508(1) & -0.&000615(9) 
 & 0.&0246(2)\\
exact & -22.&52 & -0.&0560 & 0.&73 & 0.&506 & -0.&00058(5) & 0.&02453\\
\end{tabular}
\end{table}

\begin{table}
\tabcolsep=0.1in
\caption{Computed expectation values and correlation functions from 
CPMC for a $4\times 4$ lattice with $7\uparrow\,7\downarrow$
electrons and $U=4$, 
compared with exact results. Results are shown for two different
trial wave functions $|\psi_{T1}\rangle$ and $|\psi_{T2}\rangle$.
Exact diagonalization results are from \protect\cite{diag}; numbers in
parentheses indicate either the range of values due to the
ground-state degeneracy or uncertainties in extracting numbers from a
graph. Statistical errors on the last digit of the CPMC results are in
parentheses. 
Symbols are the same as in Table III. $n_k$ is momentum distribution.}
\begin{tabular}{r@{\ \ }l r@{}l r@{}l r@{}l r@{}l r@{}l r@{}l}
 & & 
\multicolumn{2}{c}{$E_k$} & 
\multicolumn{2}{c}{$\rho(1,0)$} & 
\multicolumn{2}{c}{$\rho(2,2)$} & 
\multicolumn{2}{c}{$S(\pi,\pi)$} &
\multicolumn{2}{c}{$S_d(\pi,\pi)$} &
\multicolumn{2}{c}{$n_k(\pi/2,0)$} \\ \hline
variational&$|\psi_{T1}\rangle$ &-24.&0 & 0.&1875 & -0.&0625 & 1.&654 
 & 0.&625 & 1.&0\\
 &$|\psi_{T2}\rangle$ & -21.&88 & 0.&1706 & -0.&0602 & 4.&39 & 0.&516 
 & 0.&941\\ \hline
CPMC & $|\psi_{T1}\rangle$ & -21.&44(2) & 0.&168(1) & -0.&051(1) 
 & 2.&90(1) & 0.&432(1) & 0.&92(1)\\
 &$|\psi_{T2}\rangle$ & -21.&39(8) & 0.&168(1) & -0.&049(1) & 2.&92(2) 
 & 0.&430(1) & 0.&92(1)\\  
\hline
\multicolumn{2}{c}{exact} & -21.&39(1) & 0.&168(1) & -0.&051 
 & 2.&16(2) & 0.&425 & 0.&93(1)\\  
\end{tabular}
\end{table}

\newpage

\begin{figure}
\epsfxsize=7.5in
\epsfysize=6.9in
\centerline{\epsfbox{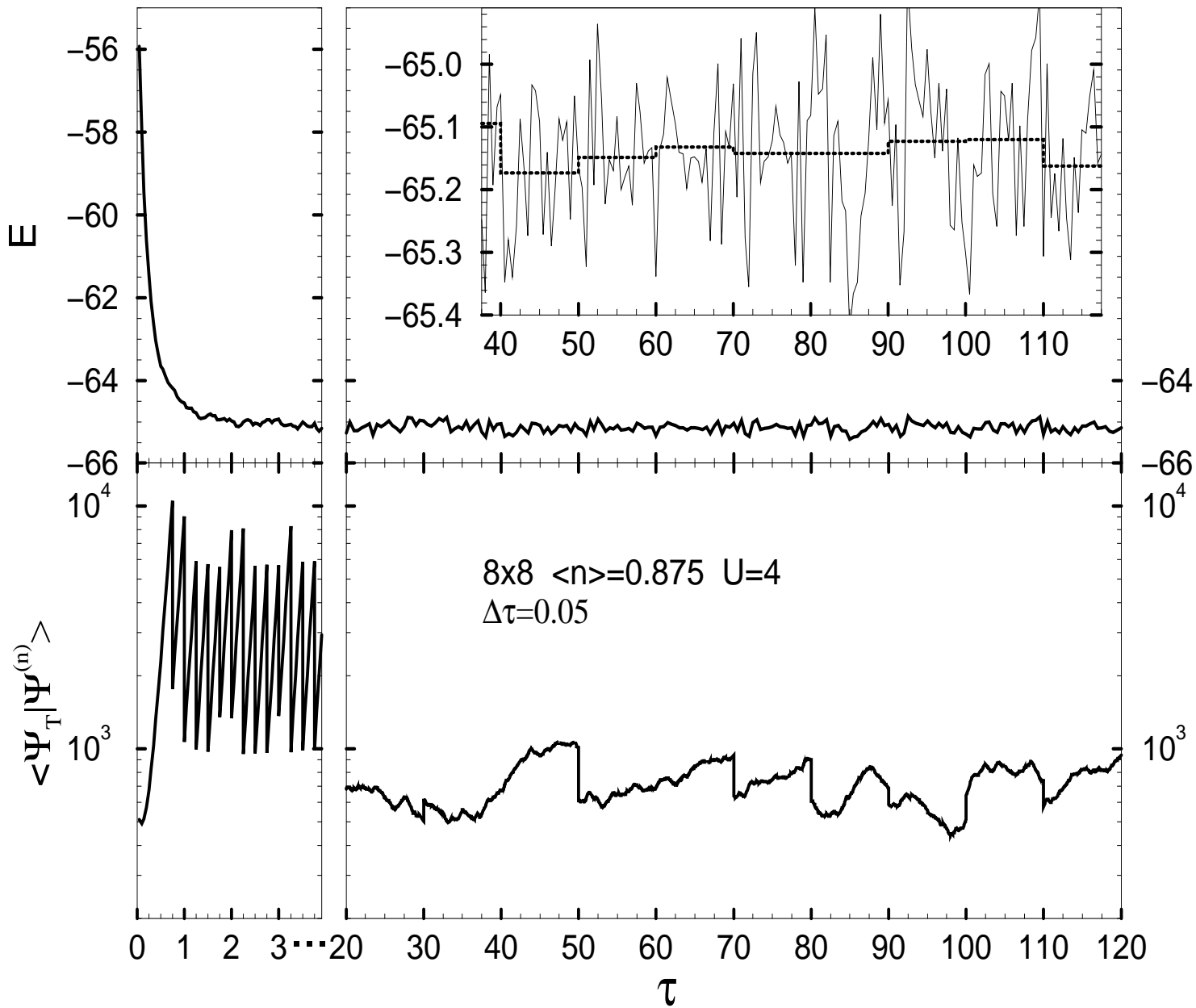}}
\vspace{0.15in}
\caption{
Stability and convergence of the CPMC method. In the lower half of the
figure is the overlap integral $\langle \psi_T|\psi^{(n)}\rangle$ as a
function of the imaginary time $\tau=n\Delta\tau$, where $n$ labels
the random walk step. In the presence of the sign problem, this
overlap integral would decay {\it exponentially\/} with $n$. In the
upper half of the figure is the corresponding estimate of the total
energy as a function of $n$. The inset is an enlarged version of the
portion between $\tau=37.5$ and $\tau=117.5$ in which the dotted line
indicates the computed energy value from blocks of length
$\tau=10$. This calculation yielded a ground-state energy of
$E_0^c=-65.135\pm 0.008$, compared to an AFQMC result
\protect\cite{Imada} of $-65.02\pm 0.06$.  }
\label{fig1}
\end{figure}

\newpage
\begin{figure}
\epsfxsize=7.5in
\epsfysize=7.2in
\centerline{\epsfbox{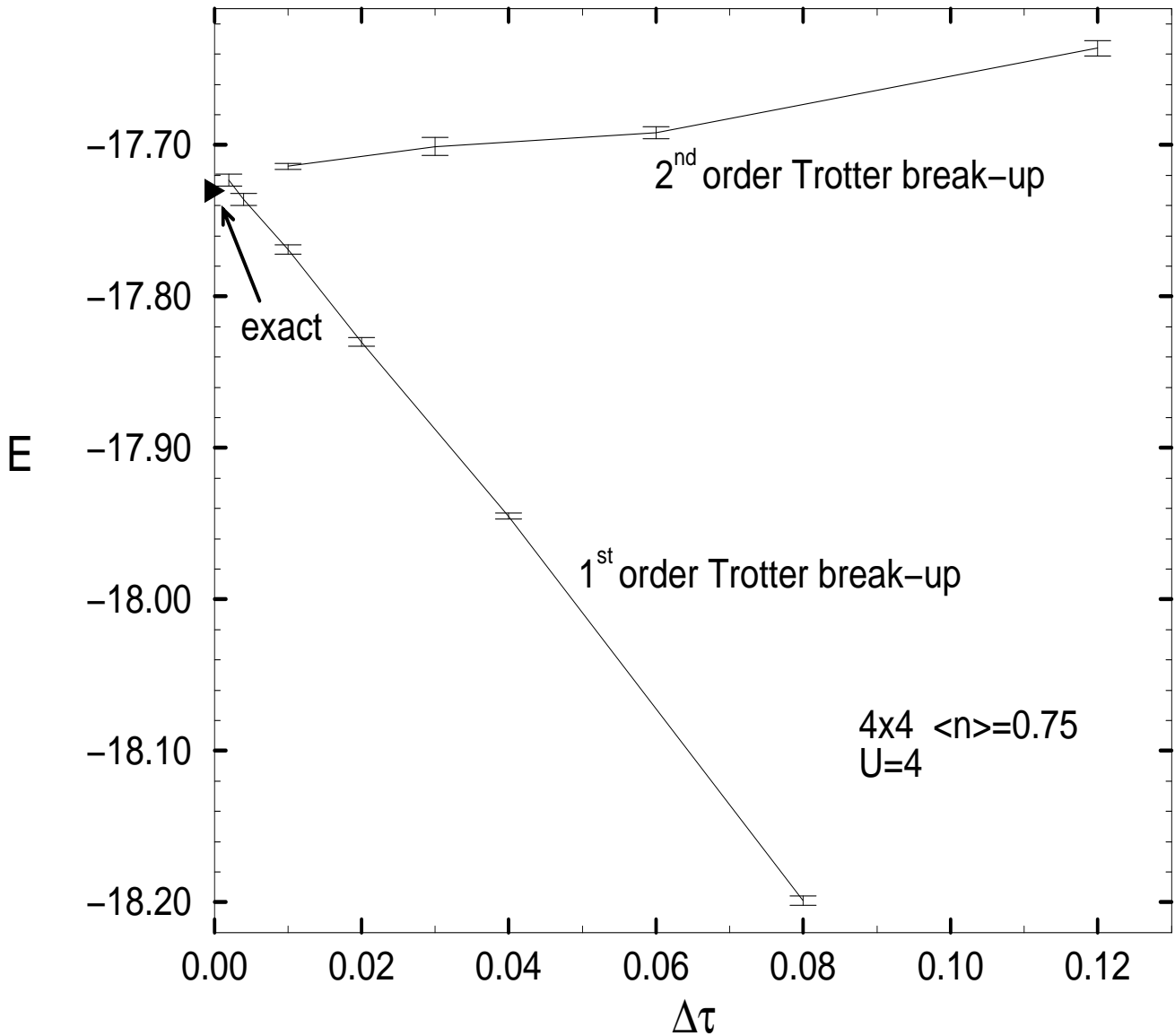}}
\vspace{0.15in}
\caption{Illustration of the Trotter approximation error. The computed
energy per site is shown as a function of the time-step $\Delta\tau$
for a $4\times 4$ lattice with $6\uparrow\,6\downarrow$ electrons and
$U=4$. The first-order Trotter approximation leads to significant
finite time-step error. With the second-order approximation,
convergence to the $\Delta\tau=0$ limit is much more rapid. The
right-triangle indicates the exact energy for this system. The
variational nature of the CPMC energies is visible. The Monte Carlo
error bars are indicated. Curves are to aid the eye.}
\label{fig2}
\end{figure}

\newpage
\begin{figure}
\epsfxsize=7.5in
\epsfysize=7.2in
\centerline{\epsfbox{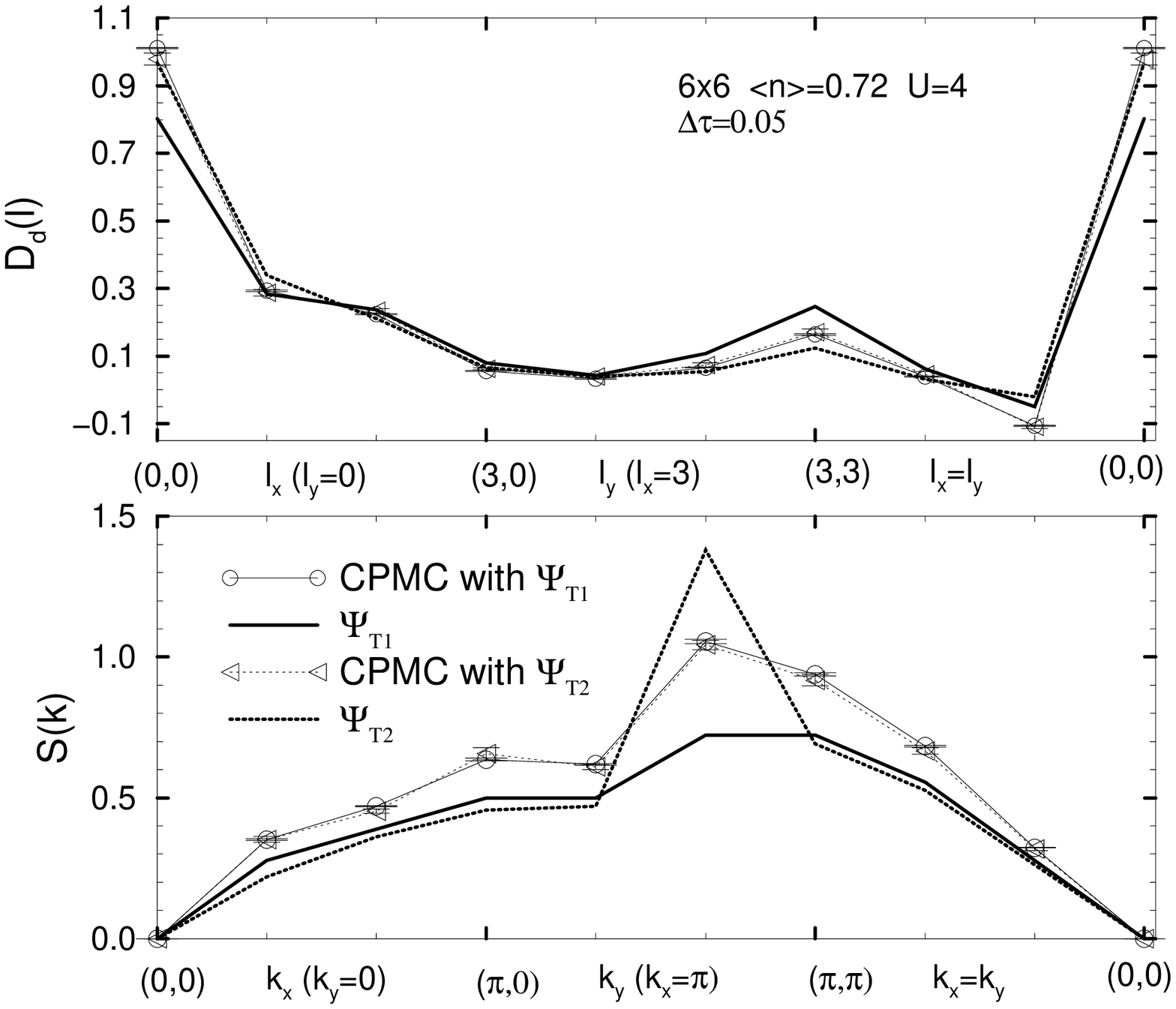}}
\vspace{0.15in}
\caption{Sensitivity of CPMC results to the choice of trial wave function
$|\Psi_T\rangle$ for some correlation functions. The upper figure is
the $d$-wave electron pairing correlation $D_{2d}({\bf l})$ and the
lower curve is the magnetic structure factor $S({\bf k})$. The
free-electron wave function $|\psi_{T1}\rangle$ and the unrestricted
Hartree-Fock wave function $|\psi_{T2}\rangle$ (with $U=4$) are used.
The corresponding mean-field results for these correlation functions
are also shown (thick lines). The computed ground-state energies from
CPMC are $-42.345(3)$ and $-42.295(16)$ (cf table II).}
\label{fig3}
\end{figure}

\end{center}


\begin{references}
\bibitem{note1} We will use the phrase ``Green's function Monte
Carlo'' in a very general sense to indicate methods that project the
ground state from a trial initial state by an importance-sampled
random walk, where wave functions are expressed in a configuration
basis and represented by a Monte Carlo sample. The difference in the
methods will be attributed to differences in the choice of the
projection operator.
\bibitem{note2} We will use the phrase ``auxiliary-field quantum 
Monte Carlo'' to refer to the class of fermion methods briefly
summarized in Section II A. We note that the adoption of this name is
relatively recent, even though the methods have been used for a number
of years, often with the generic name quantum Monte Carlo, both at
finite- and zero-temperatures. Here we use AFQMC in order to
distinguish from other QMC methods.
\bibitem{notepsiT}If $|\psi_T\rangle$ is a linear combination of 
Slater determinants, we can initialize walkers by sampling the
individual Slater determinants according to their overlap with
$|\psi_T\rangle$. Ones with negative overlaps are either avoided or
manipulated so that the form used in the initial population gives a
positive overlap.
\bibitem{O_c}
One can attempt to improve the calculation of $\langle {\cal O}
\rangle^c$ by creating 
a better importance sampling for the left-hand random walk that
both imposes the constraint with $| \psi_T \rangle$ and attempts to
incorporate the necessary correlations between the left-hand random walker
and its matching right-hand walker.  For example, we can take an
importance function $O_T^L(\phi^\prime) = | \langle \phi^\prime |
\phi_k^{(n)} \rangle | + \alpha \langle \phi^\prime | \psi_T \rangle$,
where $\alpha$ is a parameter and $|\phi_k^{(n)} \rangle$ is the
walker with which $\phi^\prime$ will be matched in the regular
population representing the right-hand wave function. While this
offers a significant improvement over the naive calculation of $\langle {\cal
O}\rangle^c$, we found it still prone to significant fluctuations.  We are
investigating other schemes to provide more accurate calculations of
the direct expectation value.
\bibitem{half_fill_imp}Care must be taken in constructing the 
importance function in the half-filled case. The definition in
(\ref{eq:O_T}) introduces an artificial ``node'' $\langle
\psi_T|\phi\rangle =0$. As $\Delta \tau\rightarrow 0$, this prevents
walkers from tunneling and incorrectly confines the random
walk. To remedy this, we use several determinants: $O_T(\phi)=\sum_j
|\langle\psi_T^j|\phi\rangle|$.  An alternative would be to shift
$O_T$ by a constant.
\end{references}
\end{document}